\pdfoutput=1

\documentclass[10pt,a4paper,twoside]{article}
\usepackage{epsfig}
\usepackage{baltlat6}
\usepackage{array}
\usepackage{here}
\begin{document}
\ \
\vspace{0.5mm}
\setcounter{page}{77}
\vspace{8mm}

\titlehead{Baltic Astronomy, vol.\,20, 77--90, 2011}

\titleb{DISKS CONTROLLING CHAOS IN A 3D DYNAMICAL MODEL \\
FOR ELLIPTICAL GALAXIES}

\begin{authorl}
\authorb{Euaggelos E. Zotos}{}
\end{authorl}

\moveright-3.2mm
\vbox{
\begin{addressl}
\addressb{}{Department of Physics,
Section of  Astrophysics, Astronomy and Mechanics,
Aristotle University of Thessaloniki,
541 24,  Thessaloniki,  Greece;\\
evzotos@astro.auth.gr}
\end{addressl}    }

\vskip-3mm

\submitb{Received: 2011 February 13; revised: March 23; accepted:
March 31}

\vskip-2mm

\begin{summary} A 3D dynamical model with a quasi-homogeneous core and
a disk component is used for the chaos control in the central parts of
elliptical galaxy.  Numerical experiments in the 2D system show a very
complicated phase plane with a large chaotic sea, considerable sticky
layers and a large number of islands, produced by secondary resonances.
When the mass of the disk increases, the chaotic regions decrease
gradually, and, finally, a new phase plane with only regular orbits
appears.  This evolution indicates that disks in elliptical galaxies
can act as the chaos controllers.  Starting from the results obtained in
the 2D system, we locate the regions in the phase space of the 3D
system, producing regular and chaotic orbits.  For this we introduce and
use a new dynamical parameter, the $S(w)$ spectrum, which proves to be
useful as a fast indicator and allows us to distinguish the regular
motion from chaos in the 3D potentials.  Other methods for detecting
chaos are also discussed.  \end{summary}


\resthead{Disks controlling chaos in elliptical galaxies}
{Euaggelos E. Zotos}

\sectionb{1}{INTRODUCTION}

Today it is believed, that dwarf elliptical galaxies seem to prefer the
densest regions of the Universe and are found abundantly in galaxy
clusters and groups (see Bingelli et al. 1987; Ferguson \& Sandage 1989;
Conselice et al. 2001).  This circumstance most likely has important
consequences for their evolution.  Moore et al.  (1998) have shown how
late-type disk galaxies, that orbit in a cluster, can loose angular
momentum by interactions with massive galaxies and, to a lesser degree,
by tidal forces induced by the cluster potential.  $N$-body simulations
performed by Mayer et al.  (2001) show that small disk galaxies, that
are close companions to a massive galaxy, will be affected likewise.  A
small disk galaxy is destabilized and develops a bar that gradually
slows down, by dynamical friction, transporting angular momentum to the
halo and to stars at larger radii.  Since the latter are being stripped,
angular momentum is lost.  Gas is funneled in towards the center by
torques exerted by the bar, where it is converted into stars, thus
forming a nucleus.  The small companion is heated by the subsequent
buckling of the bar (see Merrifield \& Kuijken 1999 and references
therein) and by bending modes of the disk, and is transformed from a
rotationally-flattened object into an anisotropic, slowly rotating
spheroidal galaxy.  The effect on a dwarf galaxy depends on its orbit
through the cluster around the massive companion.  For instance,
retrograde interactions have a much less damaging effect than prograde
ones and may even preserve some of the initial disk structure.  Thus,
these simulations allow for the existence of fast-rotating dwarfs and
for elliptical galaxies that still contain a stellar disk.

During the last decade, dwarf elliptical galaxies, such as IC 3328, IC
0783, IC 3349, NGC 4431 and IC 3468, with spiral and barred structure
have been discovered in the Virgo cluster (see Jerjen et al. 2000; De
Rijcke et al. 2001; Barazza et al. 2002; Simien \& Prugniel 2002).
Furthermore, Ryden et al. (1999) also report dwarf ellipticals with
disky isophotes in the Virgo cluster, while De Rijcke et al. (2003)
give photometric evidence for the presence of stellar disks in two
Fornax dwarf galaxies FCC 204 and FCC 288.  Moreover, there are
indications of the presence of disks in the giant elliptical
galaxies NGC 83 and NGC 2320 (see Young 2002, 2005).  The
semi-analytic simulations of Khochfar \& Burkert (2005) suggest that a
few tens of percent of all disky elliptical galaxies, could have grown
their stellar disks out of cold gas accreted from the intergalactic
medium.  A better understanding of the molecular gas in early-type
galaxies could provide concrete evidence either for or against this
disk scenario.

Taking this into account, we considered interesting to construct a 3D
dynamical model to study properties of the motion in an elliptical
galaxy with a disk.  In order to describe the motion in the region near
the center of such a system, we use the potential
\begin{equation}
V(x,y,z)=\frac{\omega ^2}{2}\left[\left(x^2+by^2+cz^2\right)-\epsilon
x\left(y^2+z^2\right)\right]\\
-\frac{M_d}{\sqrt{x^2+y^2+\left(\alpha +\sqrt{h^2+z^2}\right)^2}} \ .
\end{equation}
Potential (1) consists of two parts.  The first term stands for the
quasi-homogeneous core of the galaxy, the second term describes
the disk (see Miyamoto \& Nagai 1975).  The harmonic term of the
potential is valid only within certain distances from the center of the
galaxy.  In our case, integration of orbits was done up to distances
$R=\sqrt{x^2+y^2+z^2} \leq 1 $, hinting that up to these distances from
the galactic center, the harmonic term of Eq.  (1) is valid.  We must
point out, that the perturbation term $(-\epsilon
x\left(y^2+z^2\right))$ of the harmonic potential is an odd function
with respect to $x$.  We use this kind of perturbation term due to the
fact that it has a finite energy of escape.  Moreover, the first term of
potential (1) is a deformed galactic model, which is approximately
axisymmetric near the core but is deformed in its outer parts.  A
two-dimensional potential with this kind of perturbation was used by
Contopoulos et al.  (1987) in order to study large-scale stochasticity
in a Hamiltonian system of two degrees of freedom, which may represent
the inner parts of a deformed galactic model.  In the present paper, the
first term of potential (1), is the expansion of the potential used by
Contopoulos et al.  (1987) in a Hamiltonian system of three degrees of
freedom.  Similar potentials have been used in several papers, in order
to study axisymmetric galactic models (see Contopoulos \& Polymilis
1993; Siopis et al. 1995).

In Eq.  (1), $b$, $c$ and $\epsilon$ are parameters, $M_d$ is the mass,
$a$ and $h$ are the scale length and the scale height of the disk, while
$\omega$ is used for the consistency of the galactic units.  We use a
system of galactic units, where the unit of mass is $2.325 \times 10^7
M_\odot$, the unit of length is 1~kpc and the unit of time is $0.997748
\times 10^8$ yr.  The velocity unit is 10 km/s, while $G$ is equal to
unity.  In the above units we use the values:  $\omega$ = 10
km\,s$^{-1}$\,kpc$^{-1}$, $b$ = 1.1, $c$ = 1.5, $\epsilon$ =1.08,
$\alpha$ = 3, $h$ = 0.125, while $M_d$ is treated as a parameter.

The aim of this research is to investigate the role, played by the disk,
on the regular or chaotic nature of orbits.  Our investigation will be
focused in the following:  (i) we will try to connect the extent of the
chaotic regions with the mass of the disk; (ii) we will look for sticky
regions and secondary resonances introduced by the presence of the disk,
and (iii) we will seek if there exist only one unified chaotic region,
or different chaotic components.  For this purpose, we shall use, apart
of the classical methods, such as the Poincar\'{e} phase plane and the
Lyapunov Characteristic Exponents (LCEs) (Lichtenberg \& Lieberman
1992), some modern methods such as the $S(c)$ spectrum (Caranicolas \&
Papadopoulos 2007; Caranicolas \& Zotos 2010), the $P(f)$ indicator
(Karanis \& Vozikis 2008) and a new dynamical parameter, the $S(w)$
spectrum.

The results are based on the numerical integration of the equations of
motion
\begin{eqnarray}
\ddot{x}&=&-\frac{\partial \ V(x,y,z)}{\partial x} \ , \nonumber \\
\ddot{y}&=&-\frac{\partial \ V(x,y,z)}{\partial y} \ , \nonumber \\
\ddot{z}&=&-\frac{\partial \ V(x,y,z)}{\partial z} \ .
\end{eqnarray}
The Hamiltonian to the potential (2) reads
\begin{equation}
H=\frac{1}{2}\left(p_x^2+p_y^2+p_z^2\right)+V(x,y,z)=E \ ,
\end{equation}
where  $p_x,p_y,p_z$ are the momenta per unit mass conjugate to $x$, $y$
and $z$, while $E$ is the numerical value of the Hamiltonian.

The orbit calculations are based on the numerical integration of the
equation of motion (2), which was made using a Bulirsh-St\"oer routine
in Fortran 95, with double precision in all subroutines.  The accuracy
of the calculations was checked by the consistency of the energy
integral (3), which was conserved up to the twelfth significant figure.

The present paper is organized as follows:  in Section 2 we present an
analysis of the structure of the $x-p_x, y=0, p_y>0$ Poincar\'{e} phase
plane of the 2D system and the different families of orbits.  In the
same section a study of the evolution of the sticky regions is
presented.  Moreover, we investigate the evolution of different chaotic
components of the system.  In Section 3 we study the character of orbits
in the 3D system, using a new dynamical indicator, the $S(w)$ spectrum.
Special interest is given to the evolution of the sticky regions and the
chaotic components.  In Section 4, a discussion and the conclusions of
this research are presented.

\sectionb{2}{THE STRUCTURE OF THE $x-p_x$ PHASE PLANE:
CHAOTIC\\ COMPONENTS AND STICKY REGIONS}
\vskip1mm

In this section we will analyze the structure of the $x-p_x, y=0,
p_y>0$ Poincar\'{e} phase plane of the corresponding 2D potential, i.e.,
 we will consider orbits on the galactic plane $(z=0)$.  The
corresponding Hamiltonian is
\begin{equation}
H_2=\frac{1}{2}\left(p_x^2+p_y^2\right)+V(x,y)=E_2 \ ,
\end{equation}
where $E_2$ is the numerical value of the Hamiltonian.


\begin{figure*}[!tH]
\resizebox{\hsize}{!}{\rotatebox{0}{\includegraphics*{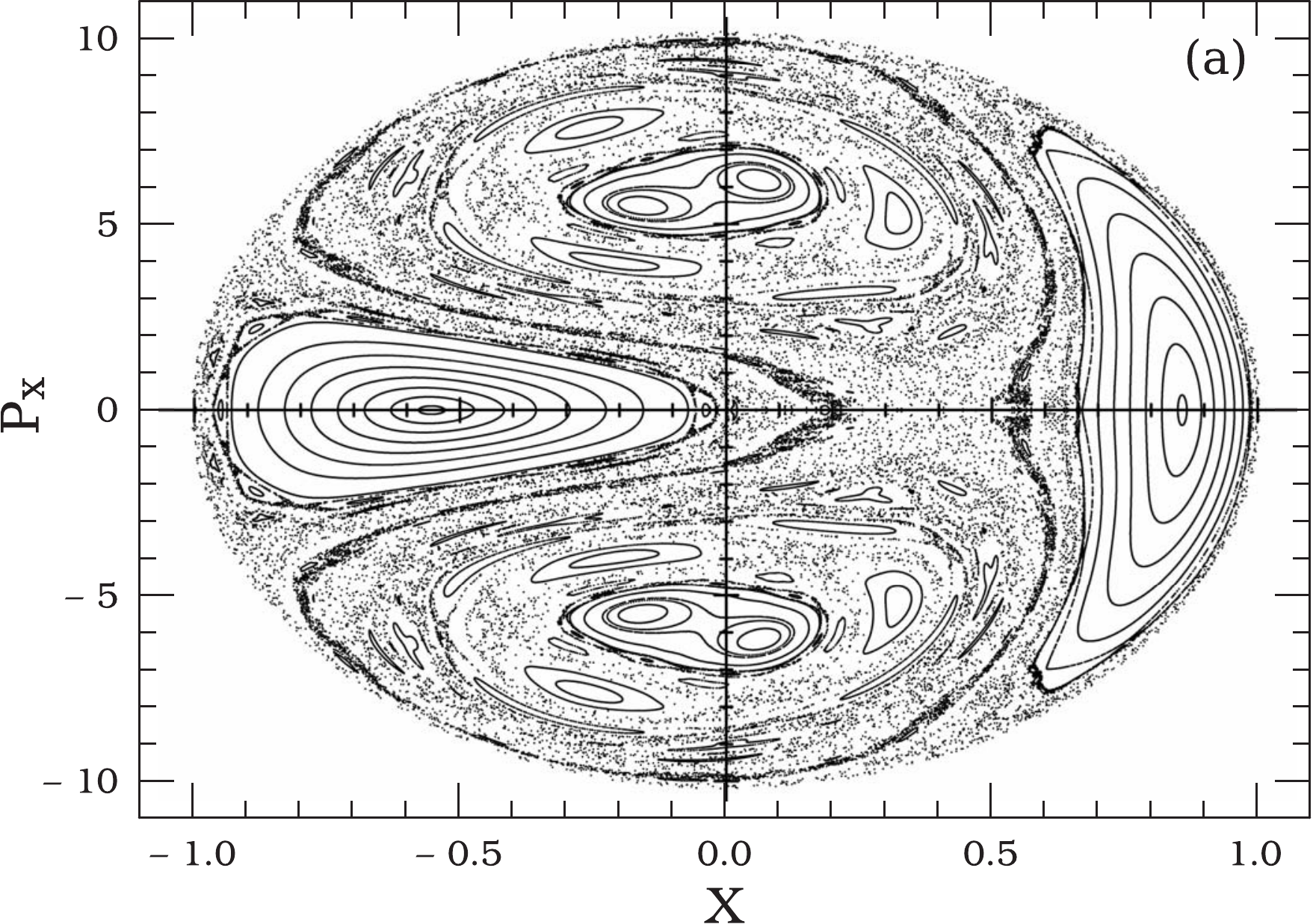}}\hspace{1cm}
                      \rotatebox{0}{\includegraphics*{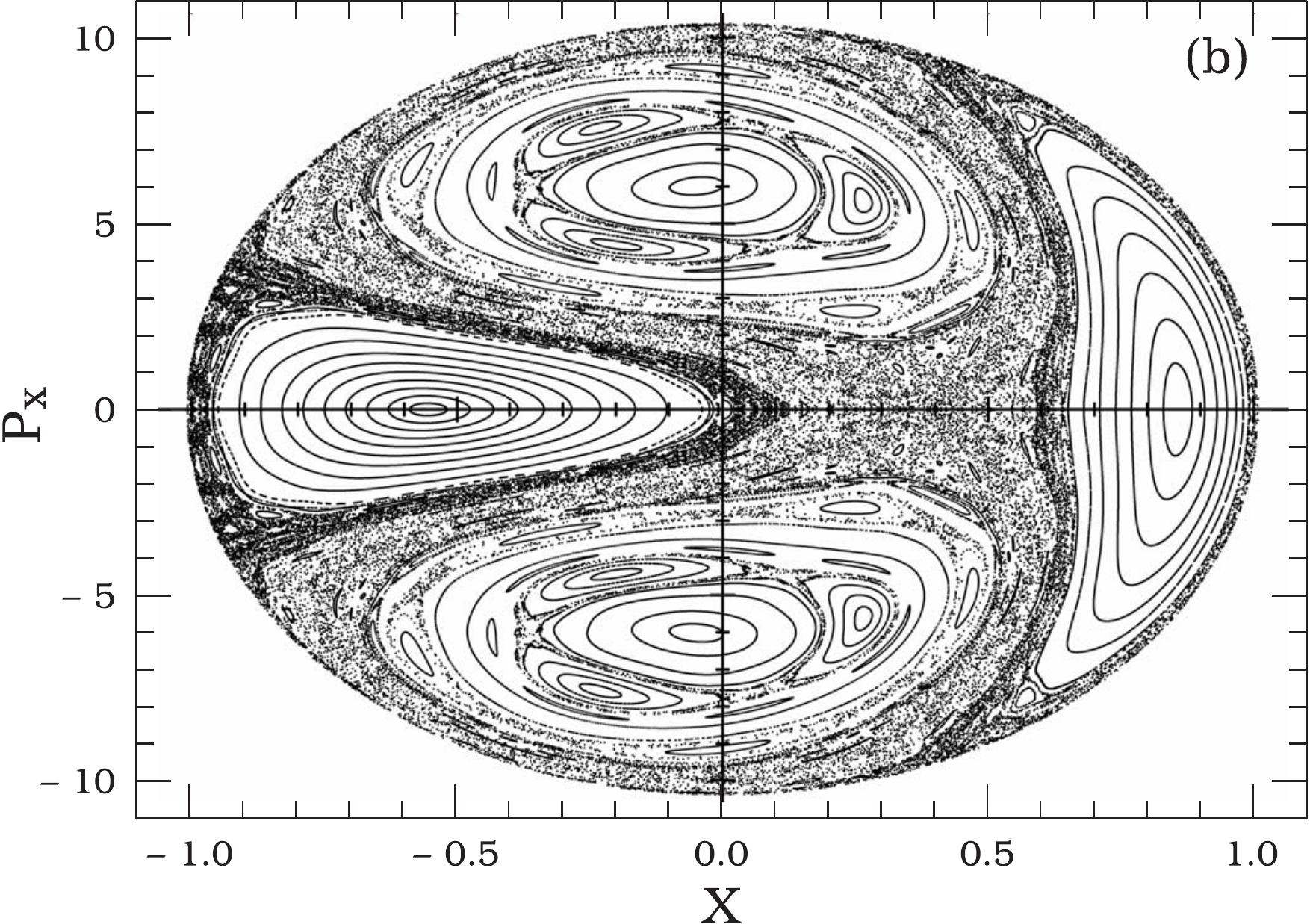}}}
\resizebox{\hsize}{!}{\rotatebox{0}{\includegraphics*{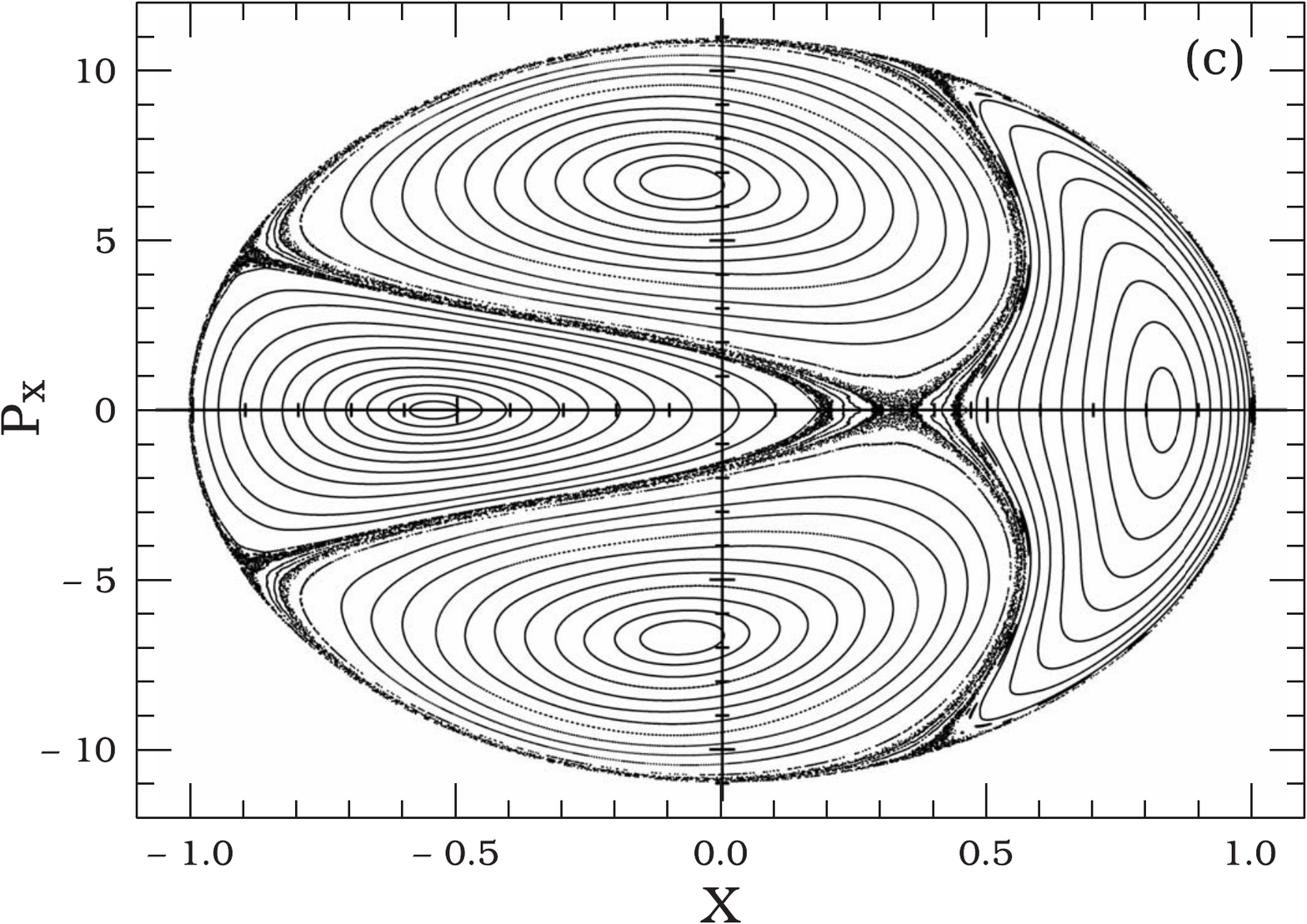}}\hspace{1cm}
                      \rotatebox{0}{\includegraphics*{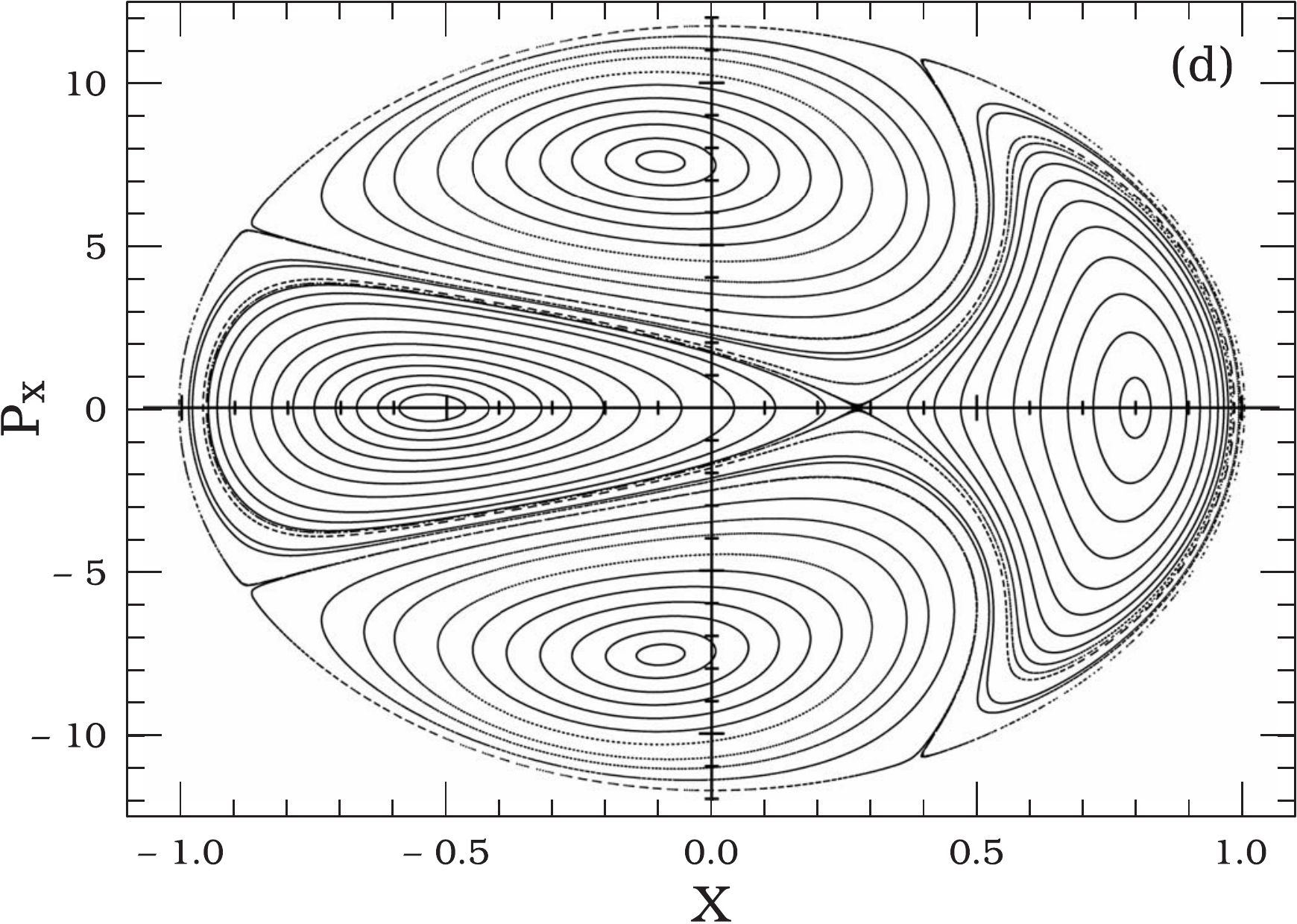}}}
\vskip 0.1cm

\captionb{1}{The $x-p_x$ phase plane.  Panel (a):  $M_d=100$, $E_2=20$;
panel (b):  $M_d=200$, $E_2=-10$; panel (c):  $M_d=600$, $E_2=-132$;
panel (d):  $M_d=1200$, $E_2=-315$.}

\end{figure*}

Figures 1 (a--d) show the $x-p_x$ phase plane of the 2D system.  Let us
start from Figure 1a, where $M_d=100$, $E_2=20$.  One observes a phase
plane with a complicated structure.  Several chaotic components seem to
exist.  There are also two main regions of regular motion.  The first
regular region, forms invariant curves around the two invariant points
on the $x$-axis, while the second region is composed of two sets of
islands intersecting the $p_x$-axis.  Both the above islands are
produced by quasi-periodic orbits characteristic of the 1:1 resonance.
A considerable regular region is occupied by two additional sets of
three islands symmetrical with respect to the $x$-axis.  Furthermore,
there are sets of smaller islands embedded in the chaotic sea and
produced by secondary resonances.  Several sticky regions seem to be
present, but we shall come to this interesting point later in this
Section.  Figure 1b shows the phase plane when $M_d=200$, $E_2=-10$.  At
a first glance, we see that the regular region has increased, while the
chaotic regions have decreased.  Note that the two invariant points near
the $p_x$-axis are now stable.  The chaotic components are present in
this case too.  The different chaotic components will be studied later
in this Section.  We must also point out, that in this case we can see
one main sticky region and some smaller secondary sticky regions as
well.  Figure 1c shows the $x-p_x$ phase plane in the case when
$M_d=600$, $E_2=-132$.  Here things are different.  Almost the entire
phase plane is regular, while only a small chaotic layer with some tiny
sticky regions is observed.  On the other hand, all periodic points are
now stable, and no secondary resonances are observed.  Figure 1d is
similar to Figure 1c, but for $M_d=1200$, $E_2=-315$.  In this case, the
entire phase plane is covered by regular orbits, while the percentage of
chaotic orbits, if any, is negligible.


\begin{figure*}[!tH]
\resizebox{\hsize}{!}{\rotatebox{0}{\includegraphics*{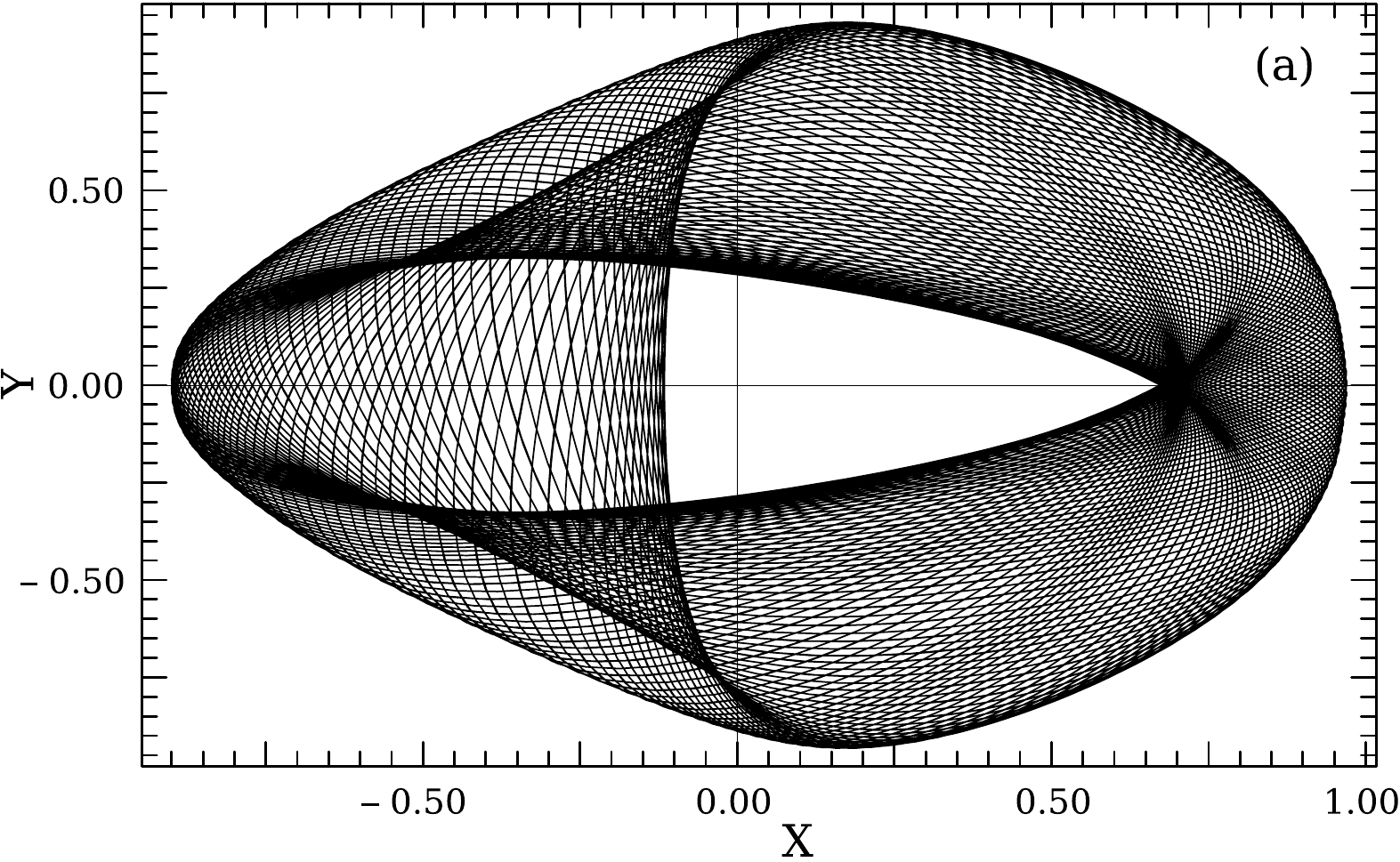}}\hspace{1cm}
                      \rotatebox{0}{\includegraphics*{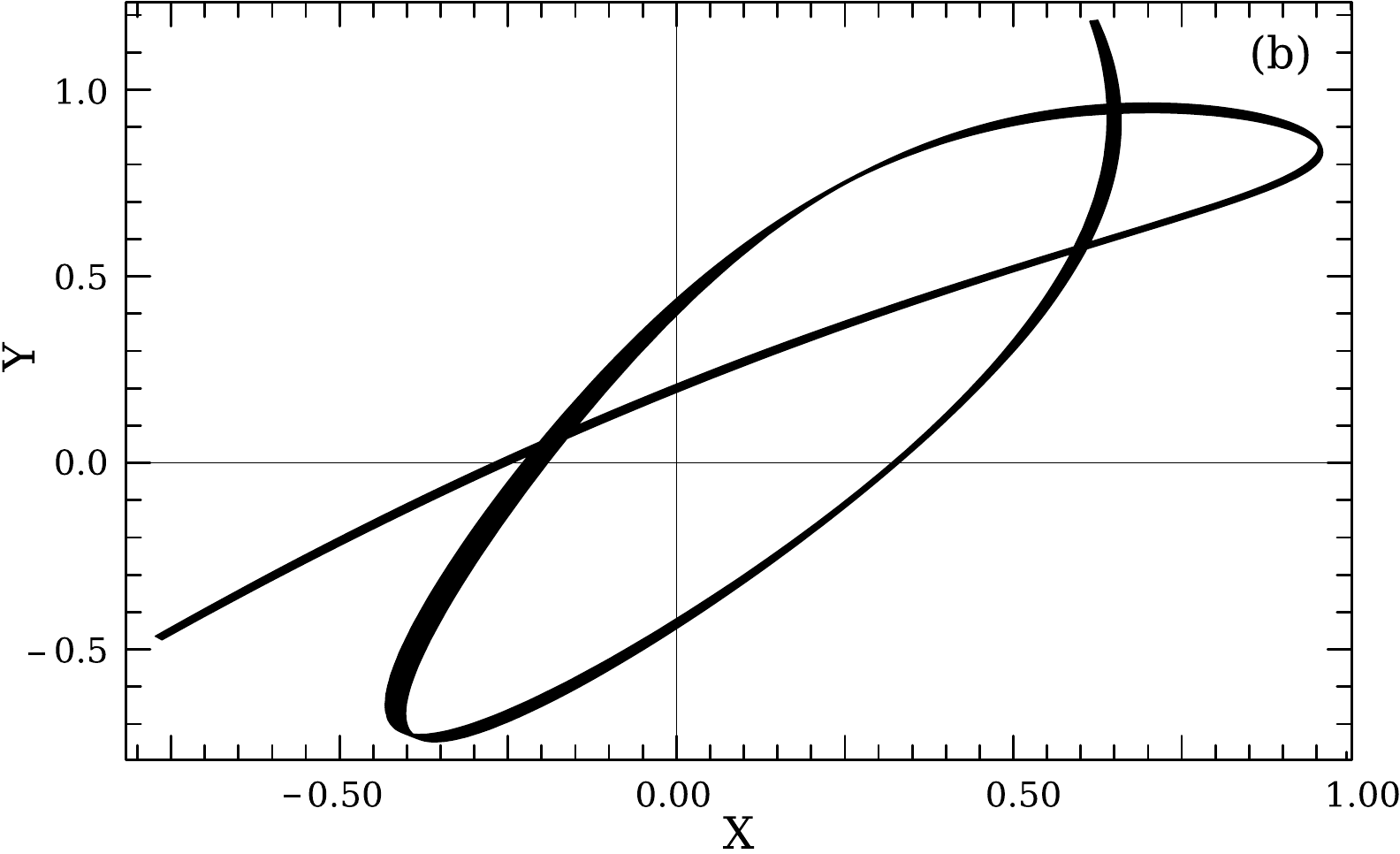}}}
\resizebox{\hsize}{!}{\rotatebox{0}{\includegraphics*{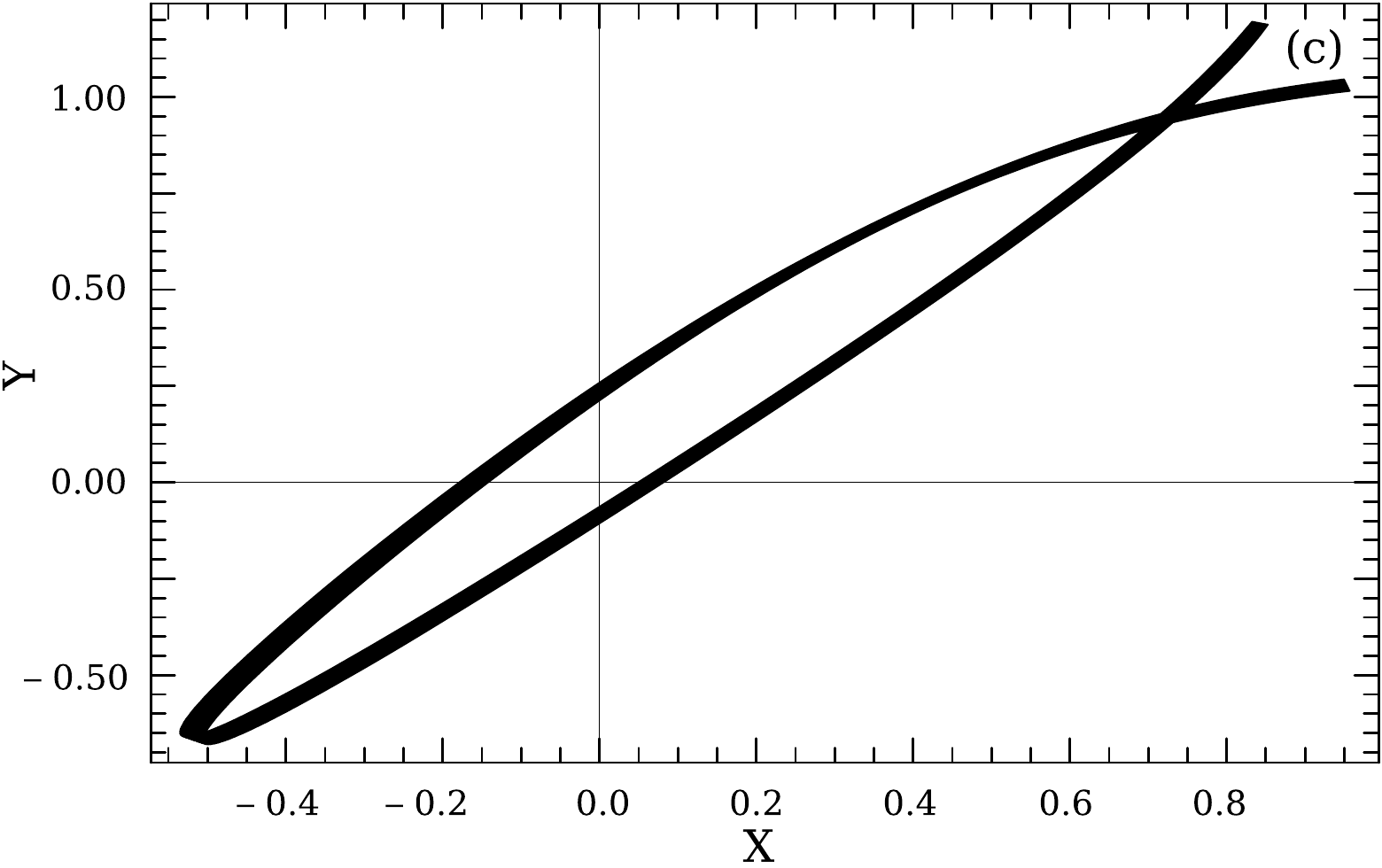}}\hspace{1cm}
                      \rotatebox{0}{\includegraphics*{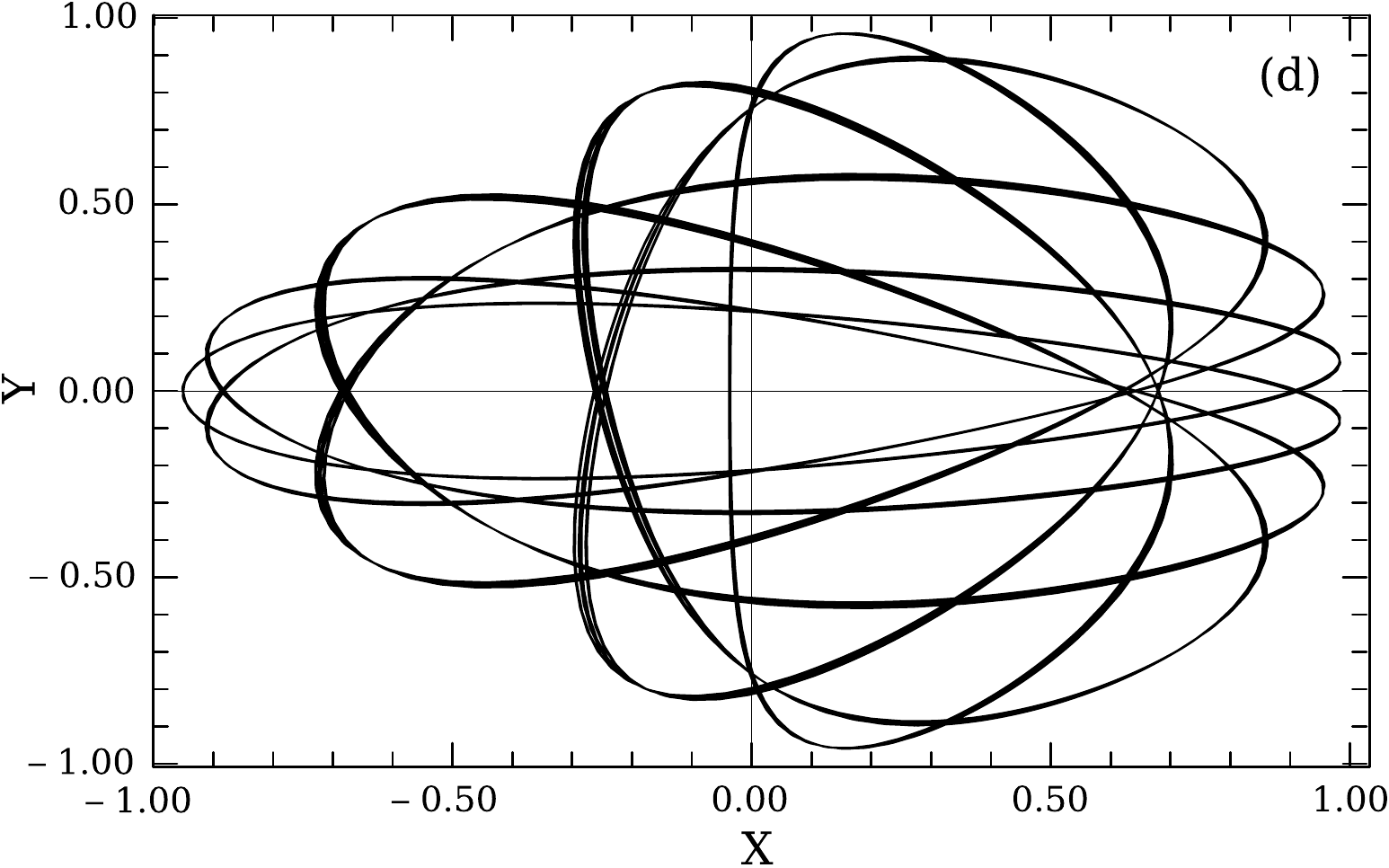}}}
\resizebox{\hsize}{!}{\rotatebox{0}{\includegraphics*{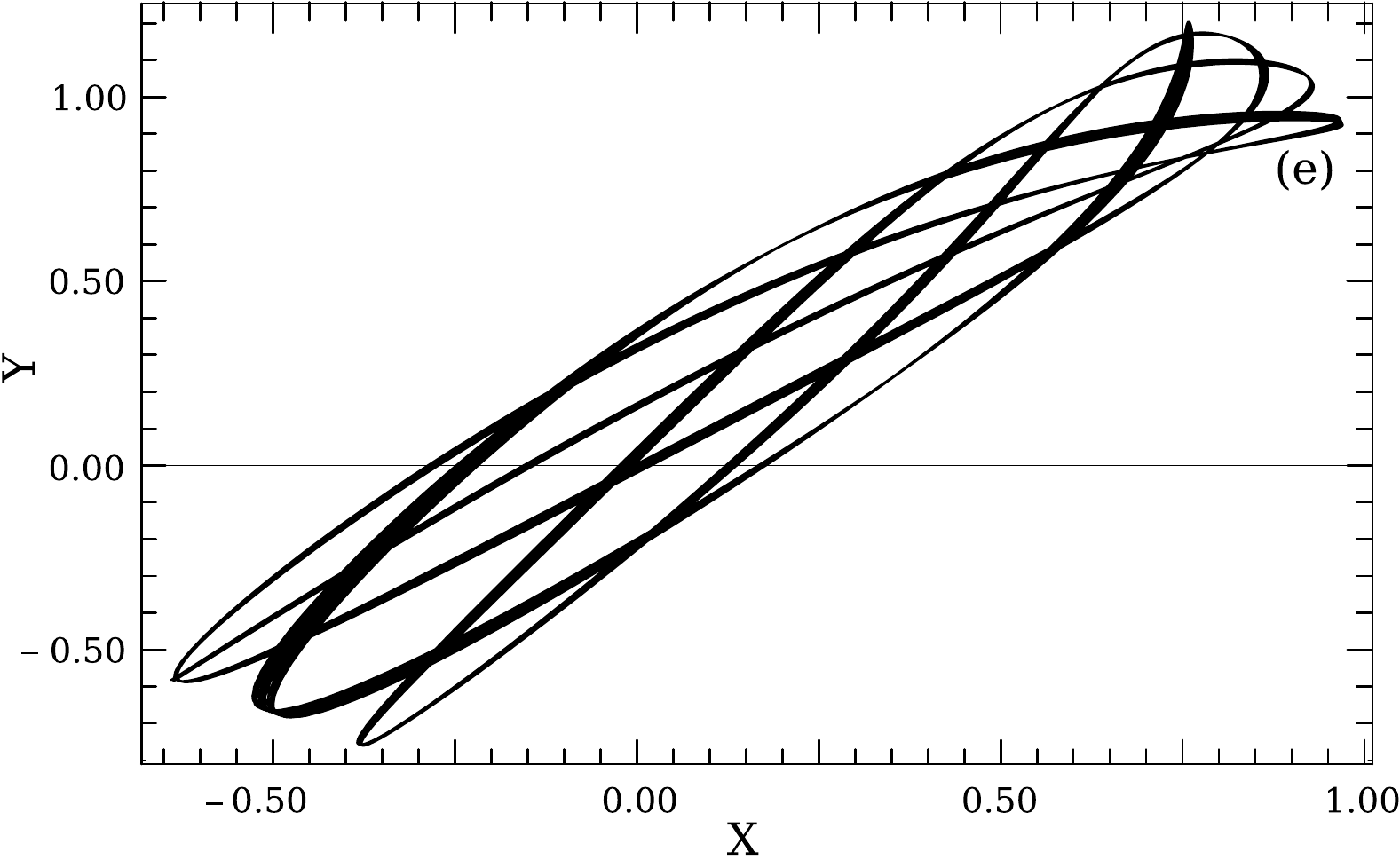}}\hspace{1cm}
                      \rotatebox{0}{\includegraphics*{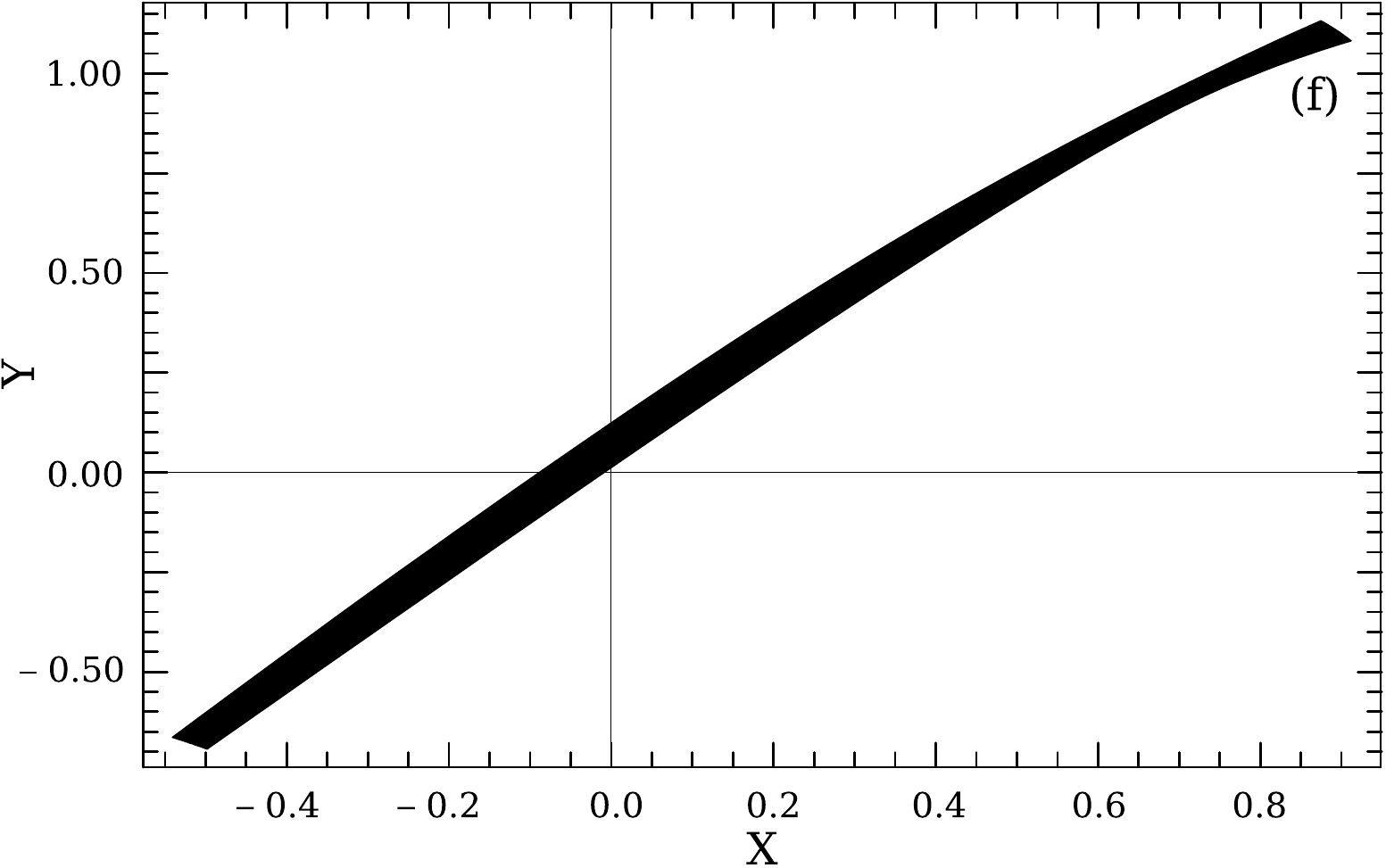}}}
\resizebox{\hsize}{!}{\rotatebox{0}{\includegraphics*{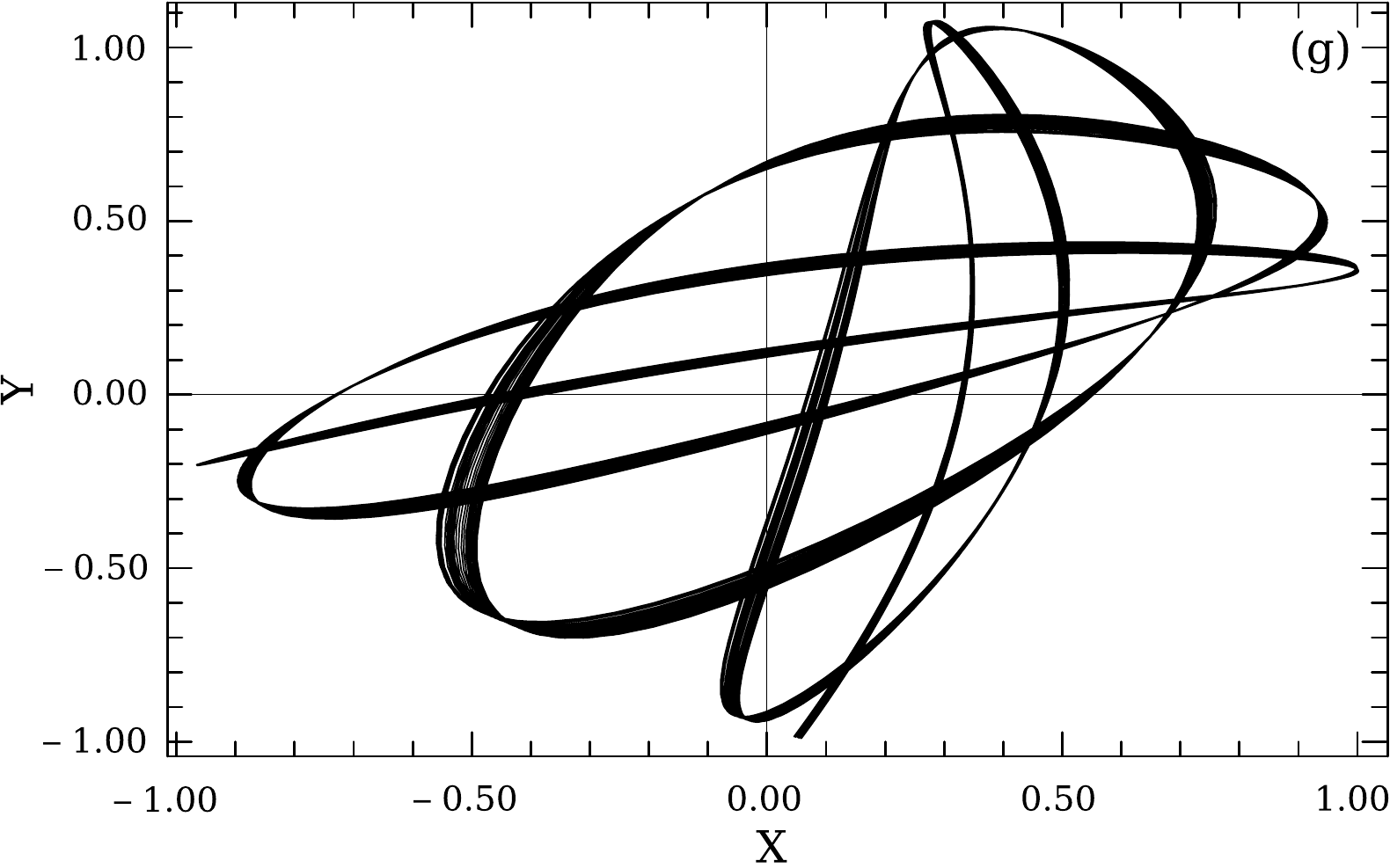}}\hspace{1cm}
                      \rotatebox{0}{\includegraphics*{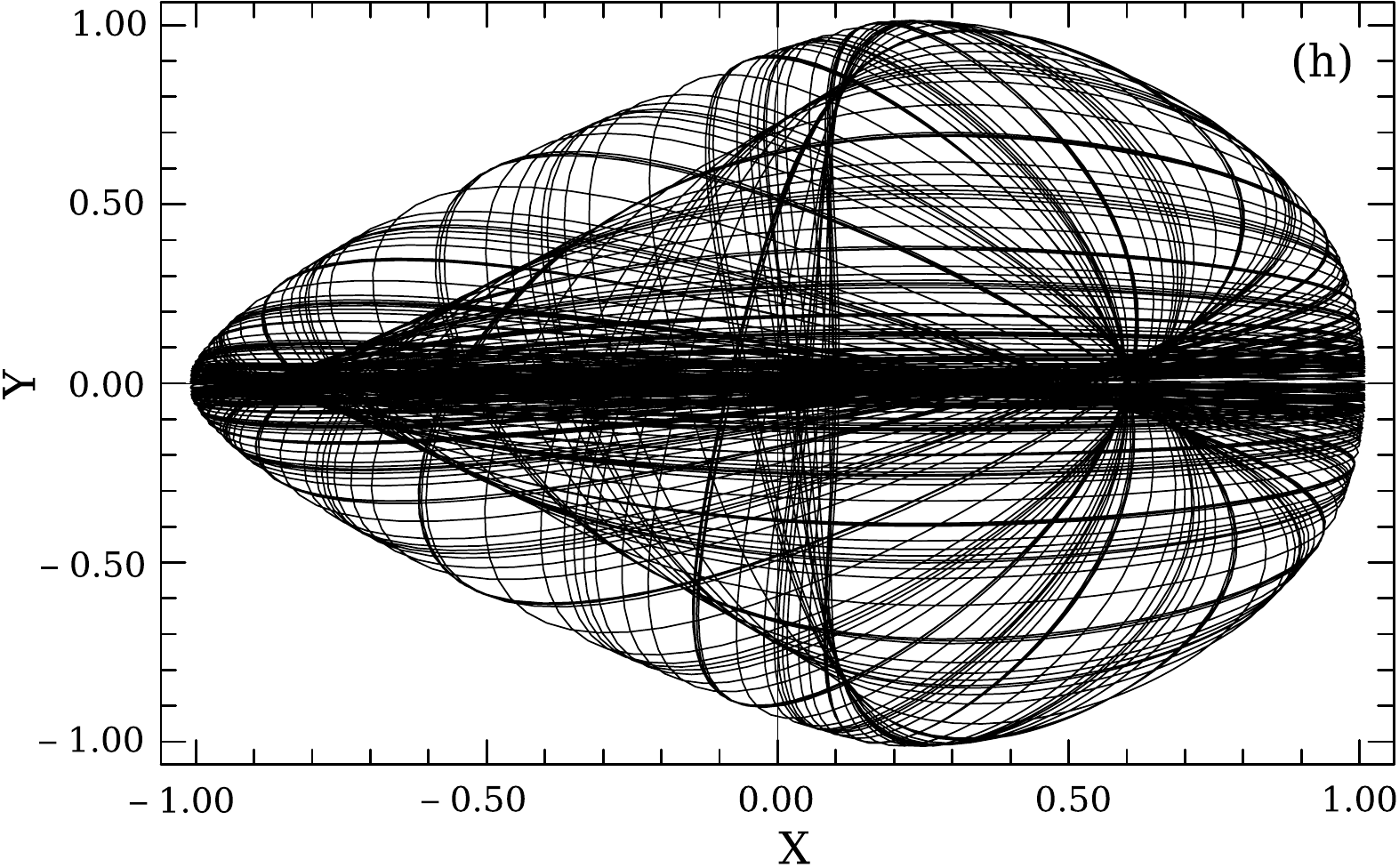}}}
\vskip 0.1cm

\captionb{2}{Orbits in the 2D potential.  The initial conditions in
panel (a):  $x_0=-0.90$, $p_{x0}=0$, panel (b):  $x_0=0.32$,
$p_{x0}=5.2$, panel (c):  $x_0=0.05$, $p_{x0}=6.2$, panel (d):
$x_0=-0.952$, $p_{x0}=0$, panel (e):  $x_0=-0.01$, $p_{x0}=4.7$, panel
(f):  $x_0=-0.01$, $p_{x0}=6$, panel (g):  $x_0=0.327$, $p_{x0}=1.16$,
panel (h):  $x_0=0.09$, $p_{x0}=0$.  In panels (a--e) $M_d=100$,
$E_2=20$, in panels (f--h) $M_d=200$, $E_2=-10$.}

\end{figure*}

Figures 2 (a--h) show several representative orbits in the 2D system.
In all orbits $y_0=0$, while the value of $p_{y0}$ is always found from
the energy integral.  The values of initial conditions and parameters
are given in the caption.  All orbits are regular except of the orbit
given in Figure 2h, which is chaotic.  The integration time for all 2D
orbits shown in Figures 2 (a--h), is 100 time units.


\begin{figure}[!tH]
\begin{center}
\resizebox{0.65\textwidth}{!}{\rotatebox{0}{\includegraphics*{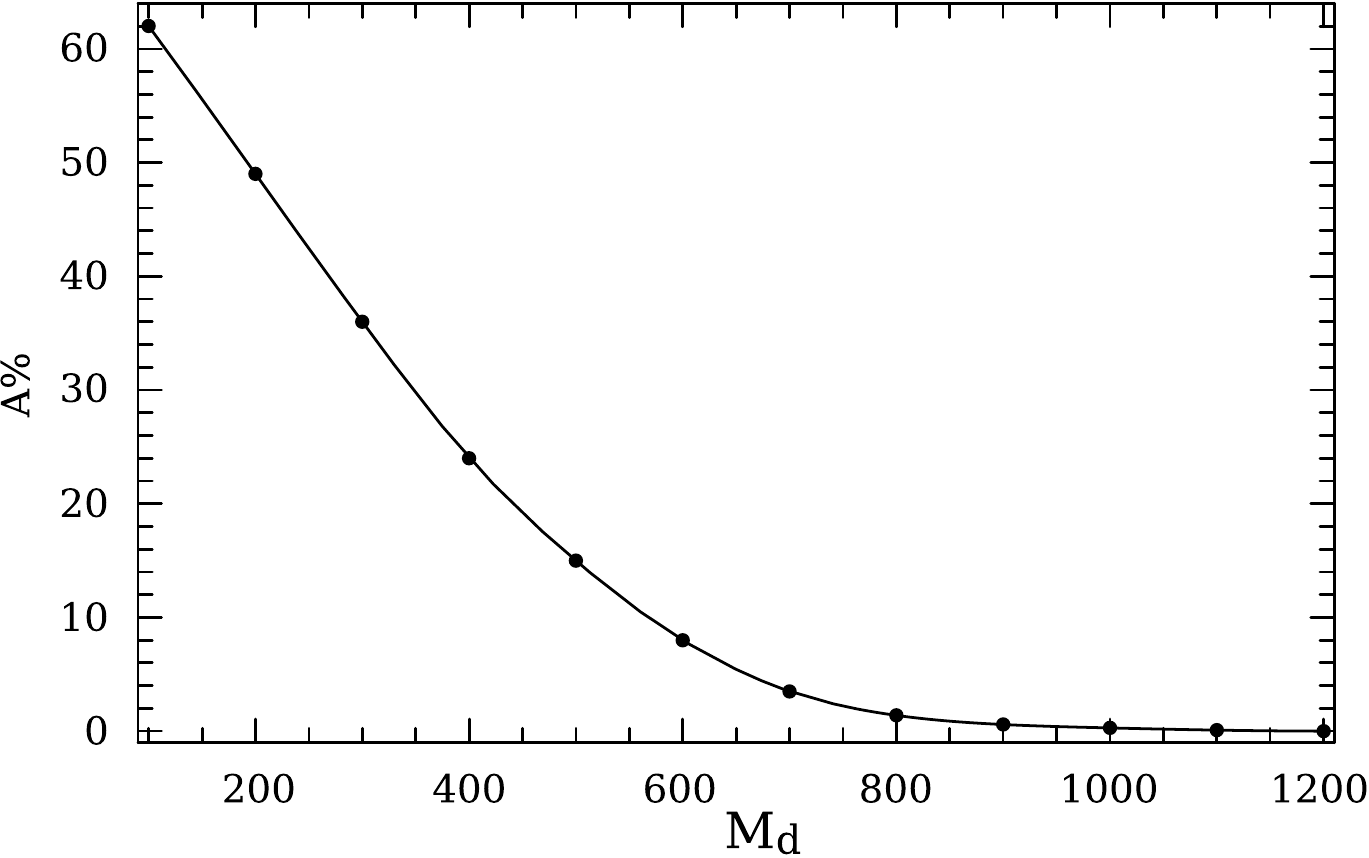}}}
\end{center}
\vspace{-5mm}
\captionc{3}{A plot of the chaotic percentage $A\%$ vs. $M_d$.}
\end{figure}

Thus we conclude, that in elliptical galaxies with massive disks in
the central regions a decrease of chaos is expected, while the
percentage of chaotic orbits is larger, when a small disk is present.
But the most important conclusion drawn from the above numerical study,
is that disks in the centers of elliptical galaxies can act as chaos
controllers.  Figure 3 shows the percentage $A\%$ of the surface of
section occupied by chaotic orbits vs.  $M_d$.  We see that $A\%$ tends
asymptotically to zero, when the mass of the disk increases.


\begin{figure}[!t]
\begin{center}
\hbox{\includegraphics[width=59.5mm,angle=0]{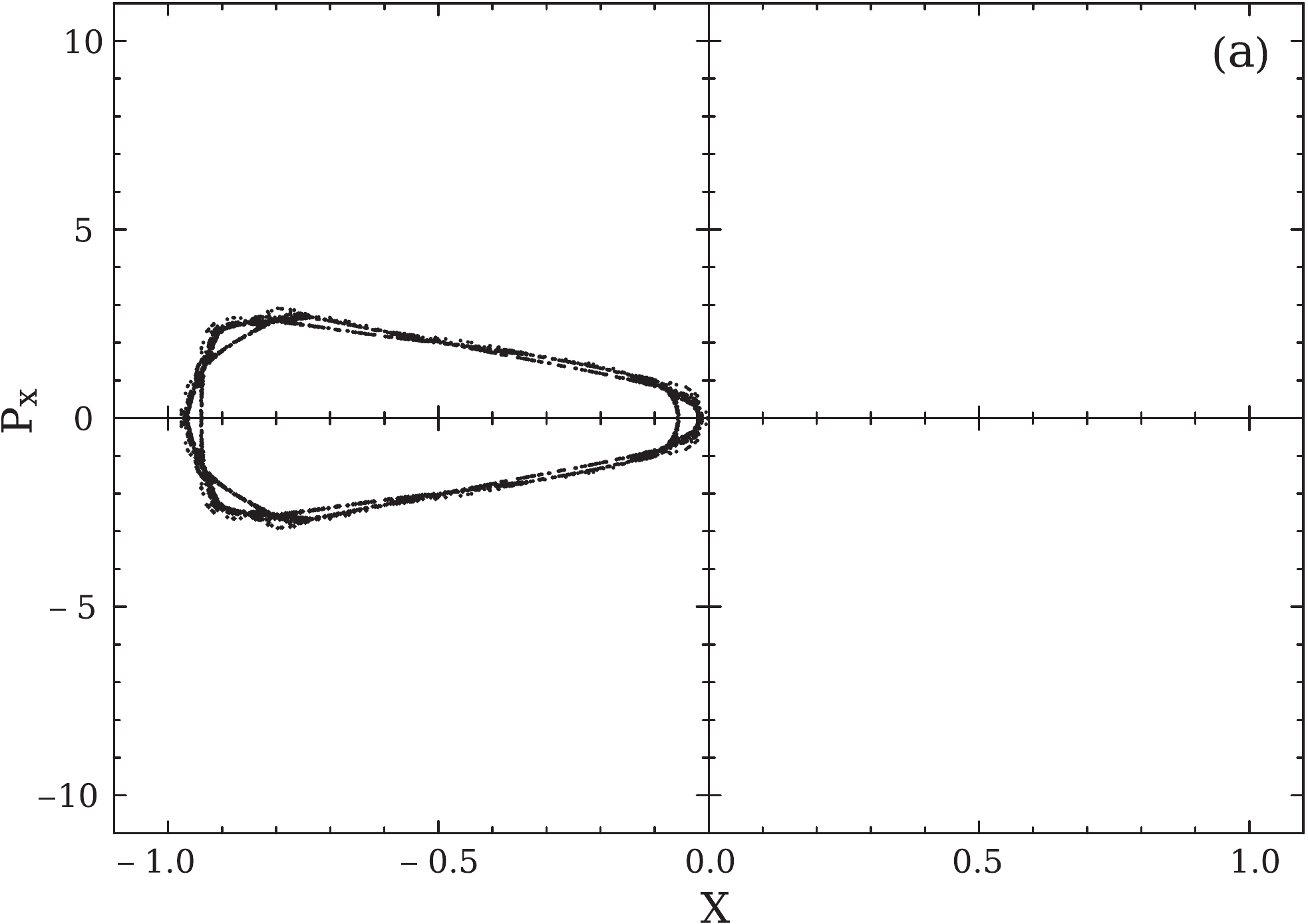}\hspace{0.3cm}
\includegraphics[width=60mm,angle=0]{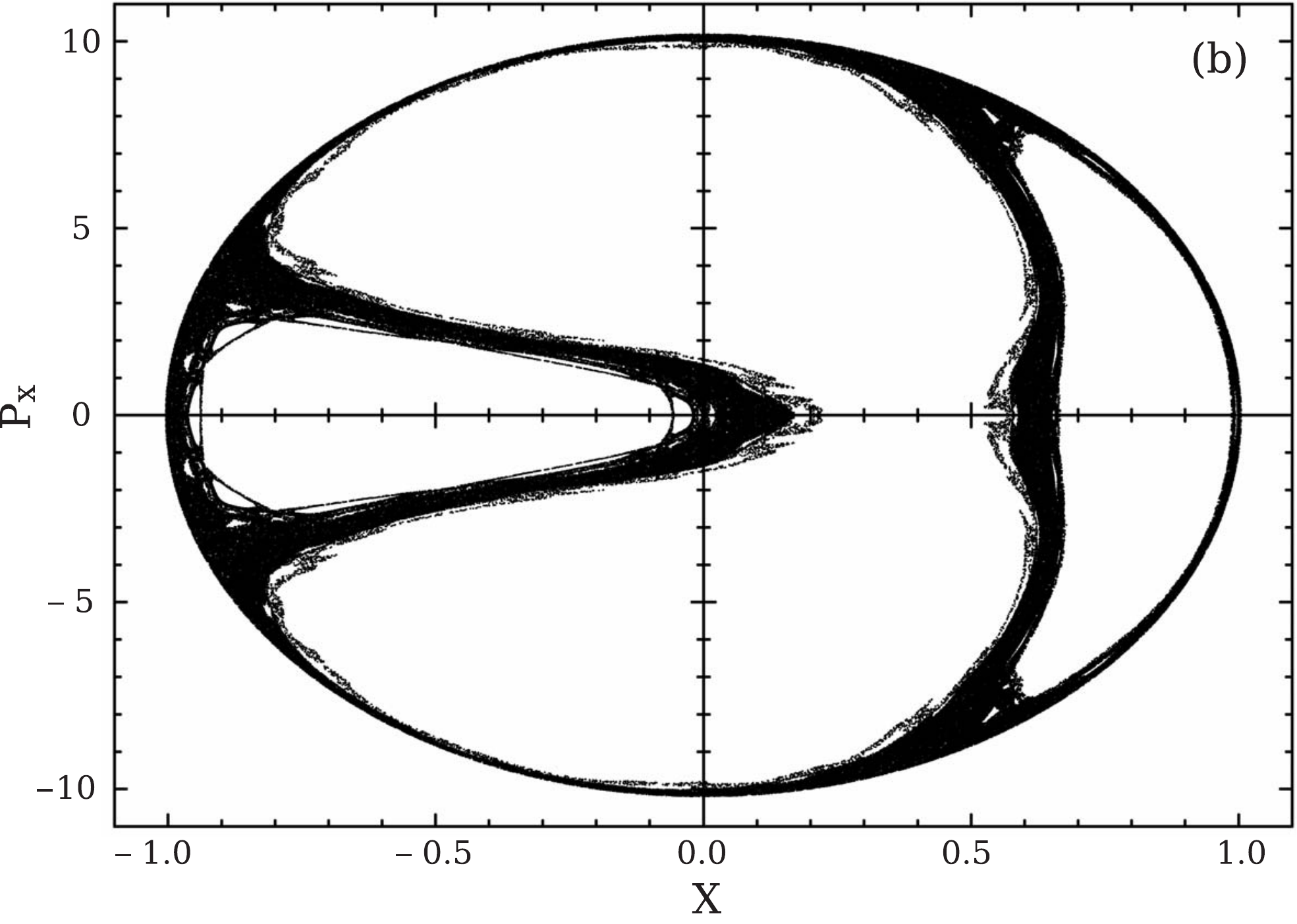}}
\vskip3mm
\hbox{\includegraphics[width=60mm,angle=0]{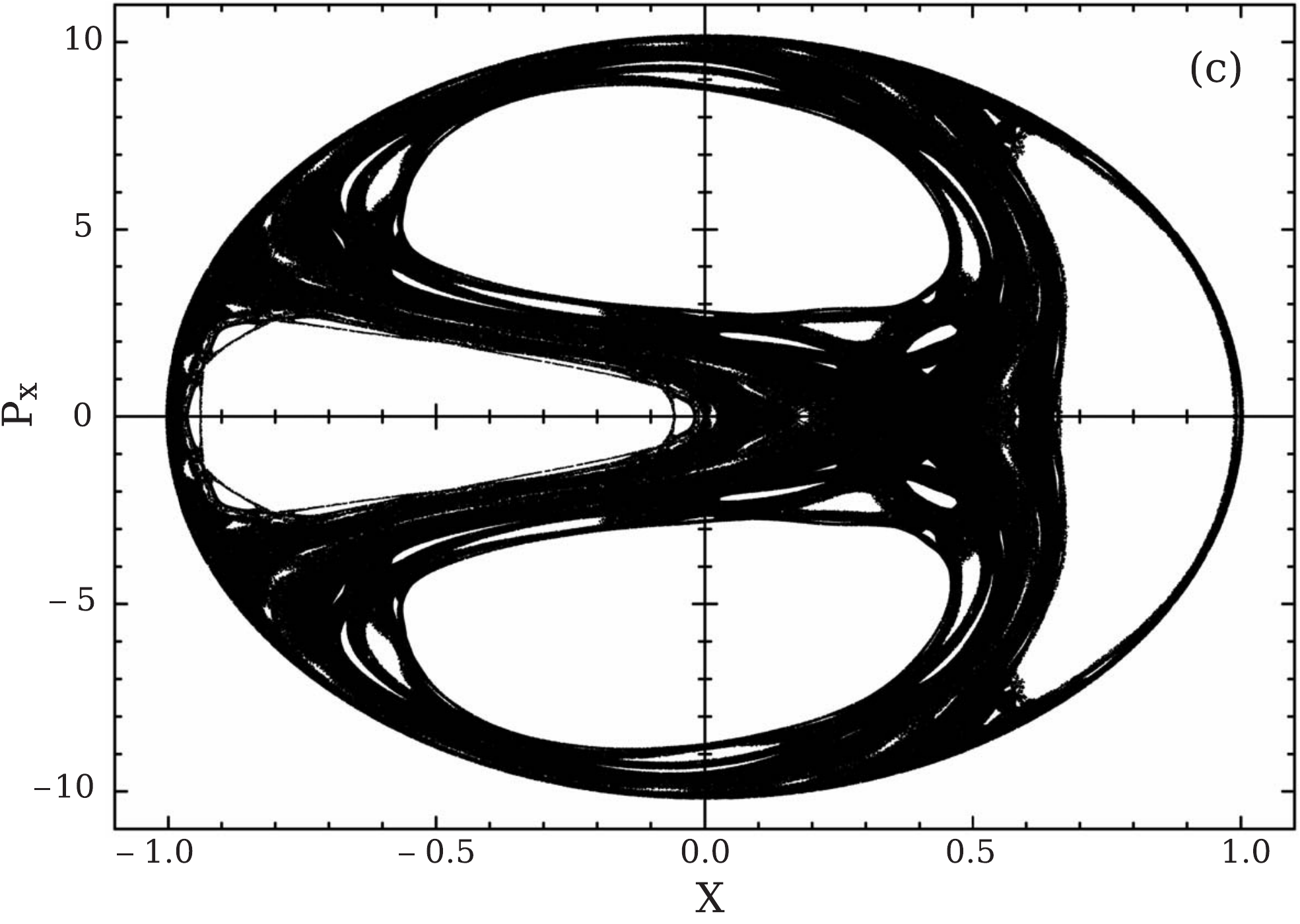}\hspace{0.3cm}
\includegraphics[width=60mm,angle=0]{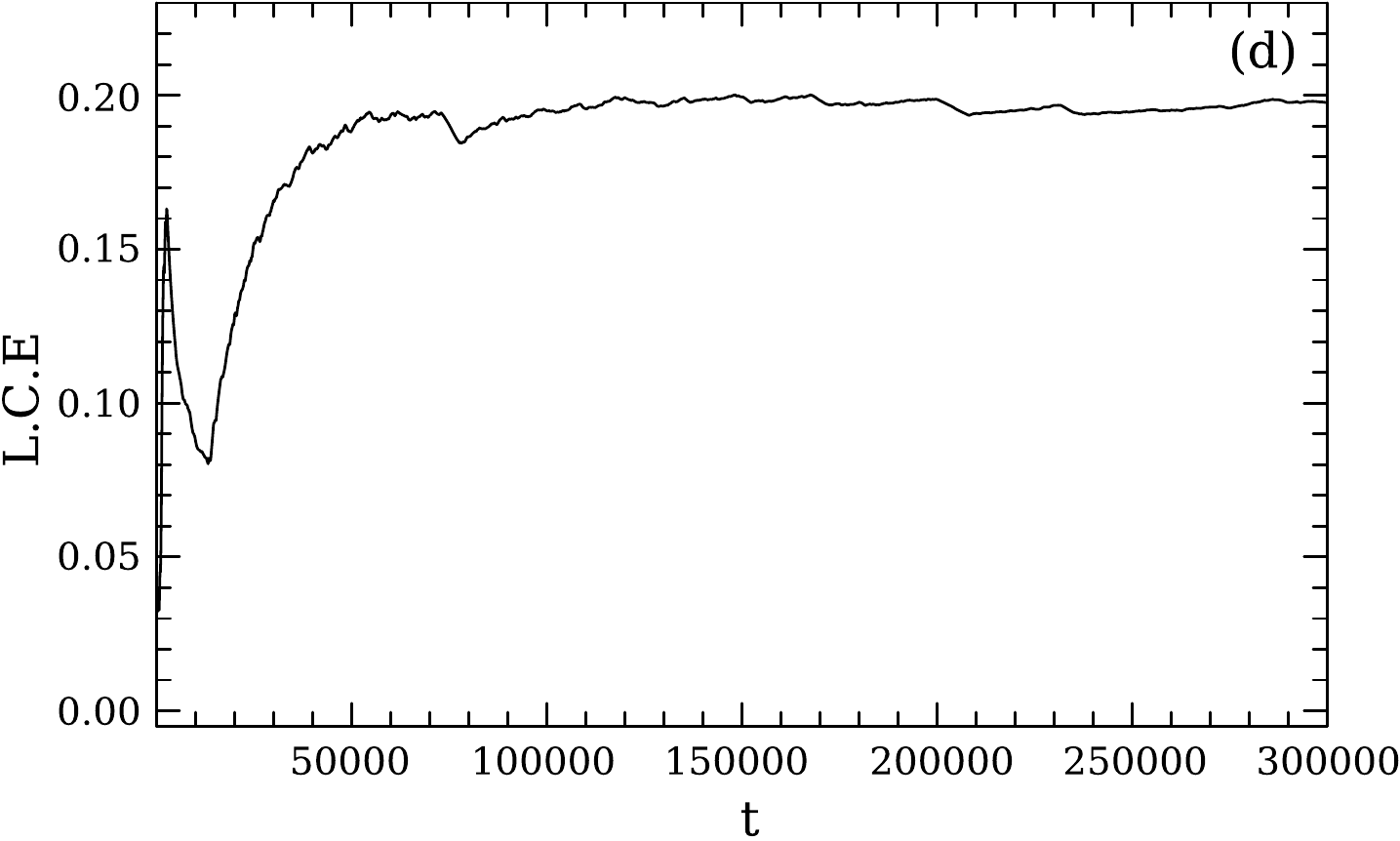}}
\end{center}
\vspace{-5mm}
\captionb{4}{Panels (a--c): evolution of a sticky orbit; panel (d):
LCE vs. time for the sticky orbit.}
\vspace{-.5mm}
\end{figure}

Now let us study the evolution of the sticky orbits of the 2D system.
We shall follow in detail the evolution of a sticky orbit with the
initial conditions:  $x_0=-0.962$, $p_{x0}=0$, the values of all other
parameters are as in Figure 1a.  The results are given in Figure 4
(a--d).  Figure 4a shows the sticky region formed in the $x-p_x$ phase
plane for a time period of about 1200 time units.  After that the test
particle leaves the above sticky region, entering to a larger sticky
region shown in Figure 4b.  There it stays to about 75\,000 time units
and then it moves to the third sticky region shown in Figure 4c, where
it stays until $3\times 10^5$ time units.  After reaching this stage,
our numerical calculations were not continued.  Our feeling is that the
evolution of the sticky orbit was completed here.  Actually we observe a
hierarchy regarding the sticky regions.  We believe that in Figure 4c we
see a chaotic component of the 2D system, not a sticky region.  Figure
4d shows a plot of the LCE of the sticky orbit for a time period of
$3\times 10^5$ time units.

\begin{figure*}[!tH]
\begin{center}
\resizebox{0.95\hsize}{!}{\rotatebox{0}{\includegraphics*{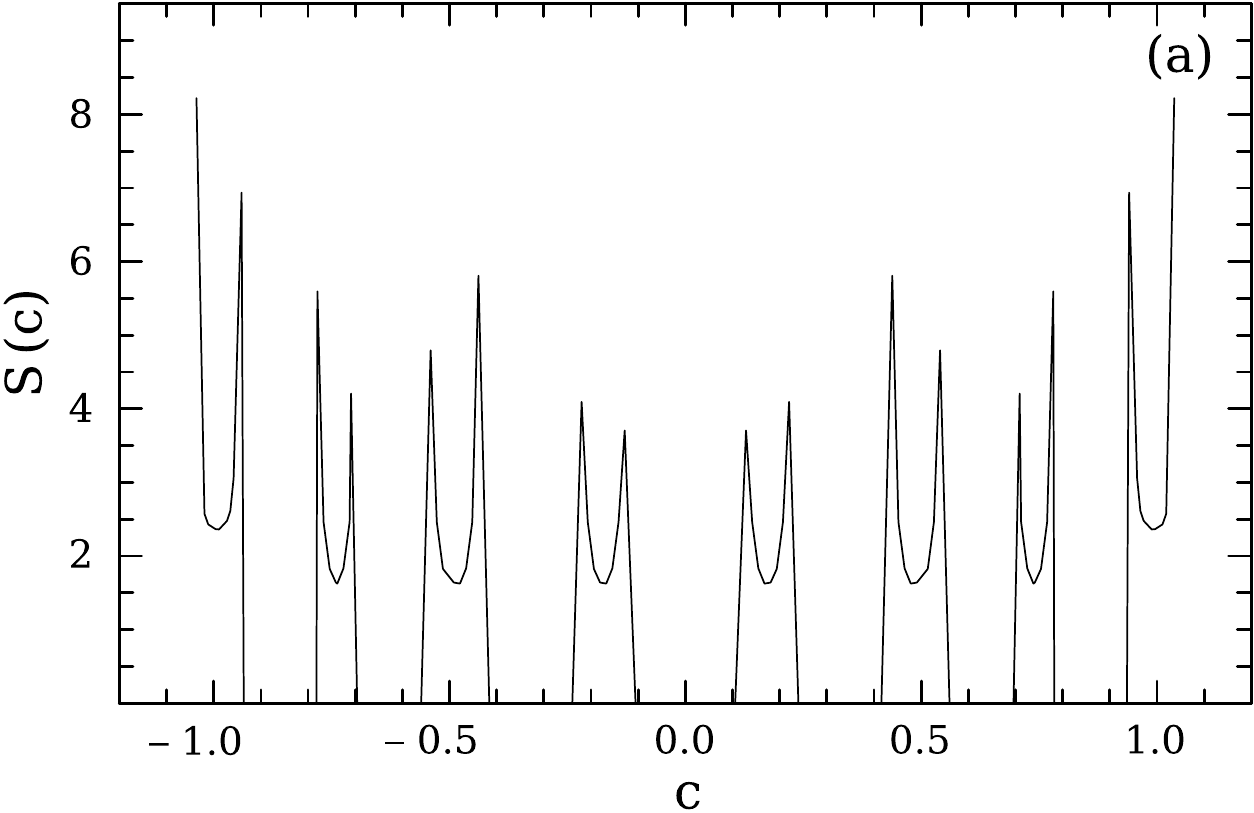}}\hspace{1cm}
                          \rotatebox{0}{\includegraphics*{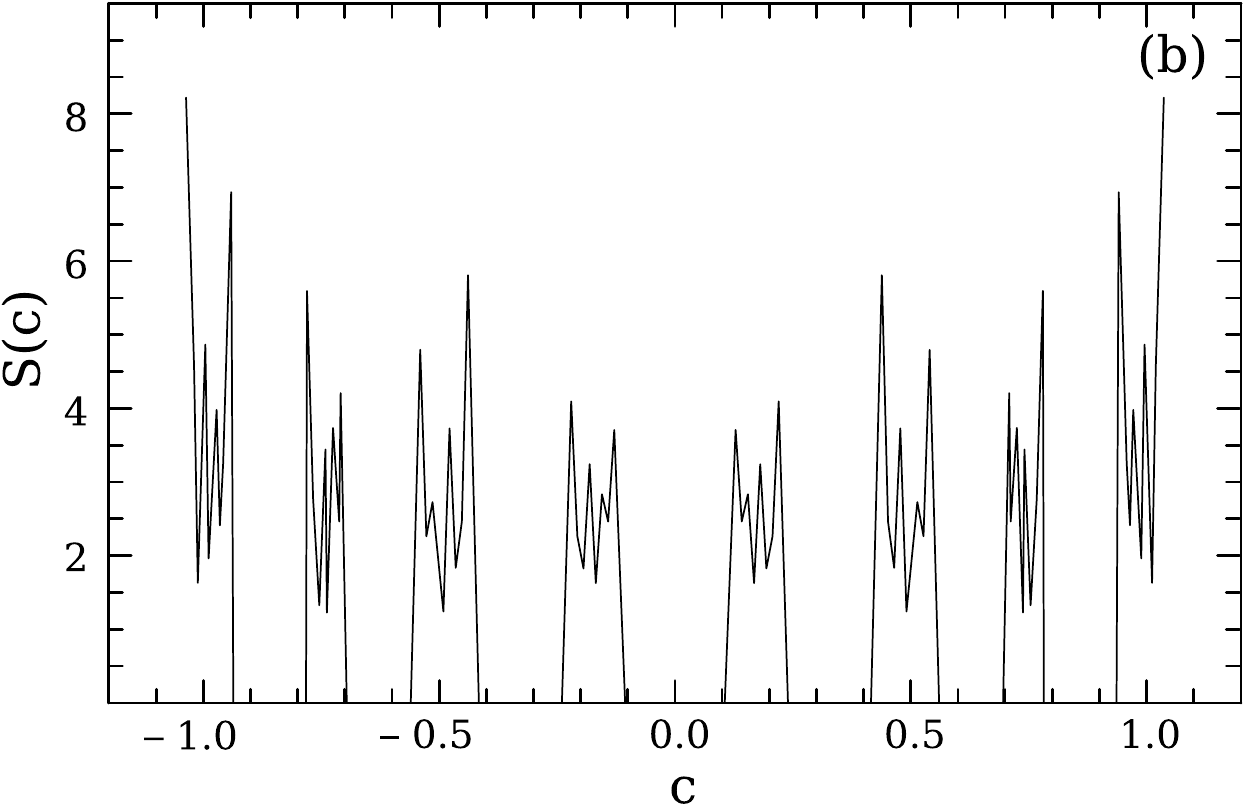}}}
\resizebox{0.95\hsize}{!}{\rotatebox{0}{\includegraphics*{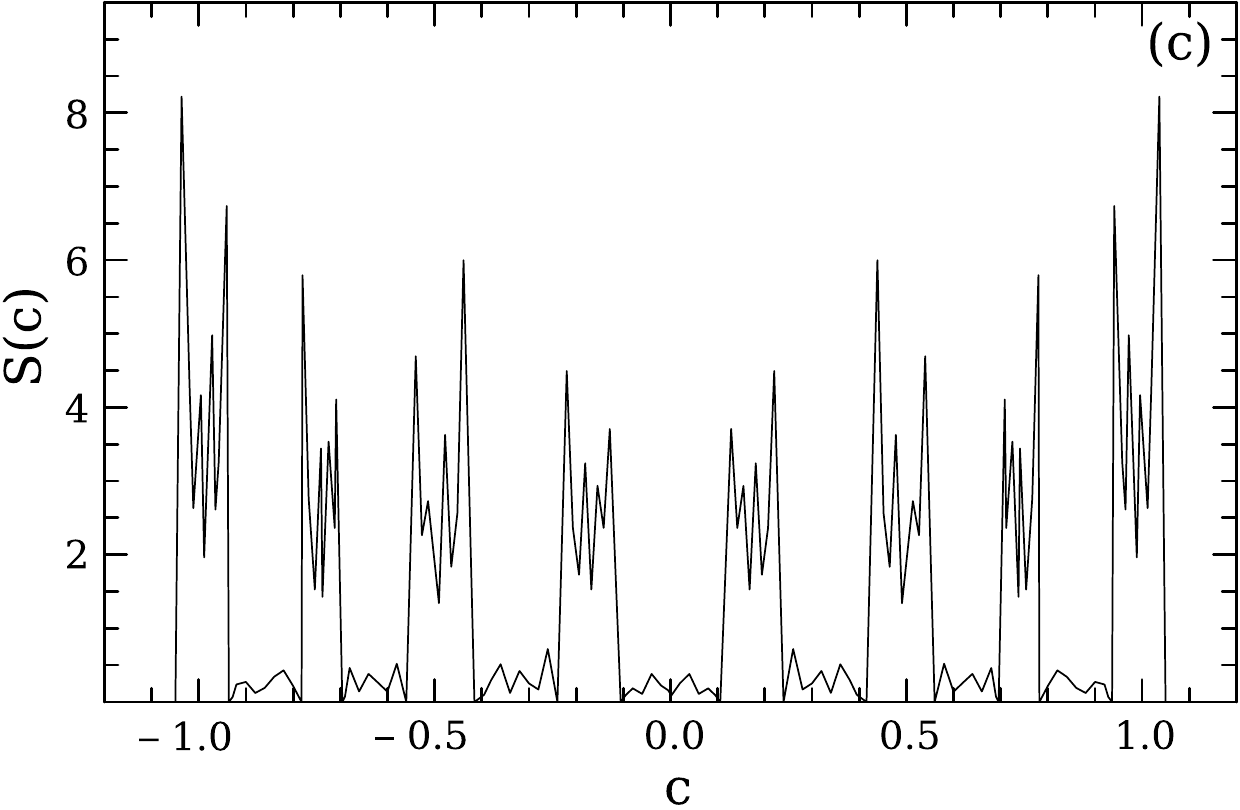}}\hspace{1cm}
                          \rotatebox{0}{\includegraphics*{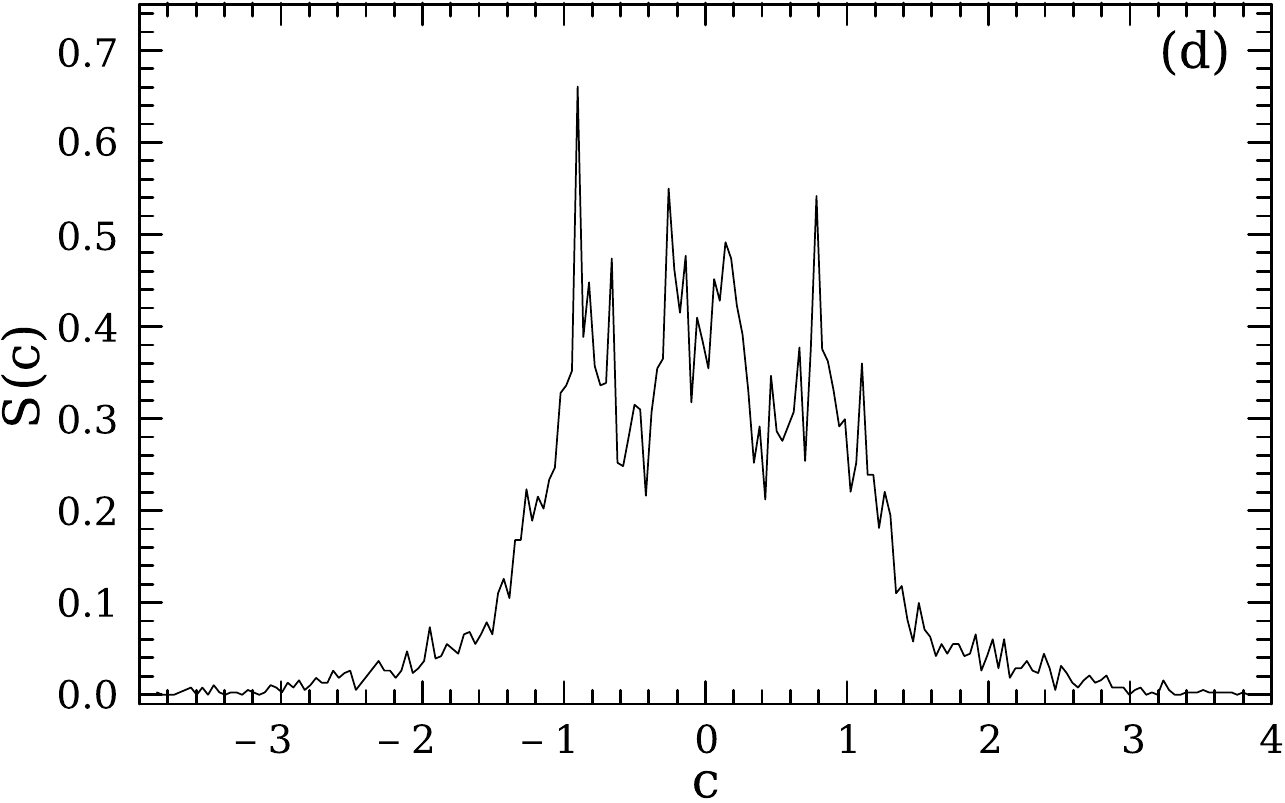}}}
\end{center}
\captionb{5}{Panel (a):  the $S(c)$ spectrum for an orbit producing the set
of eight small islands, shown in Fig.\,1a; panels (b--d):  evolution of
the $S(c)$ spectrum of a sticky orbit.  Details are given in the
text.}

\end{figure*}

A better view of the sticky orbit evolution can be seen using the $S(c)$
spectrum.  Figure 5a shows the $S(c)$ spectrum for an orbit producing a
set of eight small islands, shown in Figure 1a.  The initial conditions
are:  $x_0=-0.952$, $y_0=p_{x0}=0$, and the values for all other
parameters are as in Figure 1a.  As expected, we observe eight well
defined $U$-type spectra.  The motion is regular.  Figure 5b shows the
$S(c)$ spectrum for the sticky orbit.  Here, we can see again eight
spectra, each corresponding to an island.  The basic difference between
Figures 5a and 5b is that in Figure 5b we observe a large number of
asymmetric peaks.  Those additional peaks indicate the sticky motion.
It is well known, that in the dynamical systems of two degrees of
freedom, sticky orbits are the orbits which stay for long time periods
near the last Kolmogorov-Arnold-Moser (hereafter KAM) torus, before they
escape to the surrounding chaotic sea (see Karanis \& Caranicolas 2002).

According to the Kolmogorov-Arnold-Moser theorem, most orbits
lie on tori in the phase space. However, near all unstable periodic points,
there is some degree of stochasticity. This is best seen on the surface
of section, on which the successive intersections of orbits passing
close to the unstable periodic points, do not lie, in general, on closed
invariant curves, but fill stochastically a certain defined area. These
orbits are called stochastic, or chaotic, or semiergodic. The phenomenon
of the onset of large scale chaoticity, has been studied in more detail
in recent years. It was found, that the invariant curves that separate
chaotic regions in the neighborhood of two unstable periodic orbits are
destroyed as the energy goes beyond a critical value. Thus, we have
communication between the two chaotic regions. The critical value of the
energy occurs when the last KAM curve (or the last KAM torus in the
phase plane) separating the two resonant regions is destroyed.

Now let us go to the evolution of the
sticky orbit, using the $S(c)$ spectrum.  During the first sticky
period, which is about 1100 time units, in Figure 5b one observes eight
separate complicated spectra.  In Figure 5c the time is 1500 time units,
and the eight spectra are very similar to those seen in Figure 5b.  Note
that here the eight spectra are connected.  This indicates that the test
particle has left the first sticky region in order to continue its
wandering in a larger sticky region.  Figure 5d, shows the $S(c)$
spectrum for 70\,000 time units.  We see that the spectrum tends to take
characteristics of a chaotic spectrum.  Note that the time intervals of
the sticky periods, obtained by the evolution of the $S(c)$ spectrum,
are very close to those obtained by the formation of the $x-p_x$ phase
planes, shown in Figures 4 (a--d).

\begin{figure*}[!tH]
\begin{center}
\resizebox{1.0\hsize}{!}{\rotatebox{0}{\includegraphics*{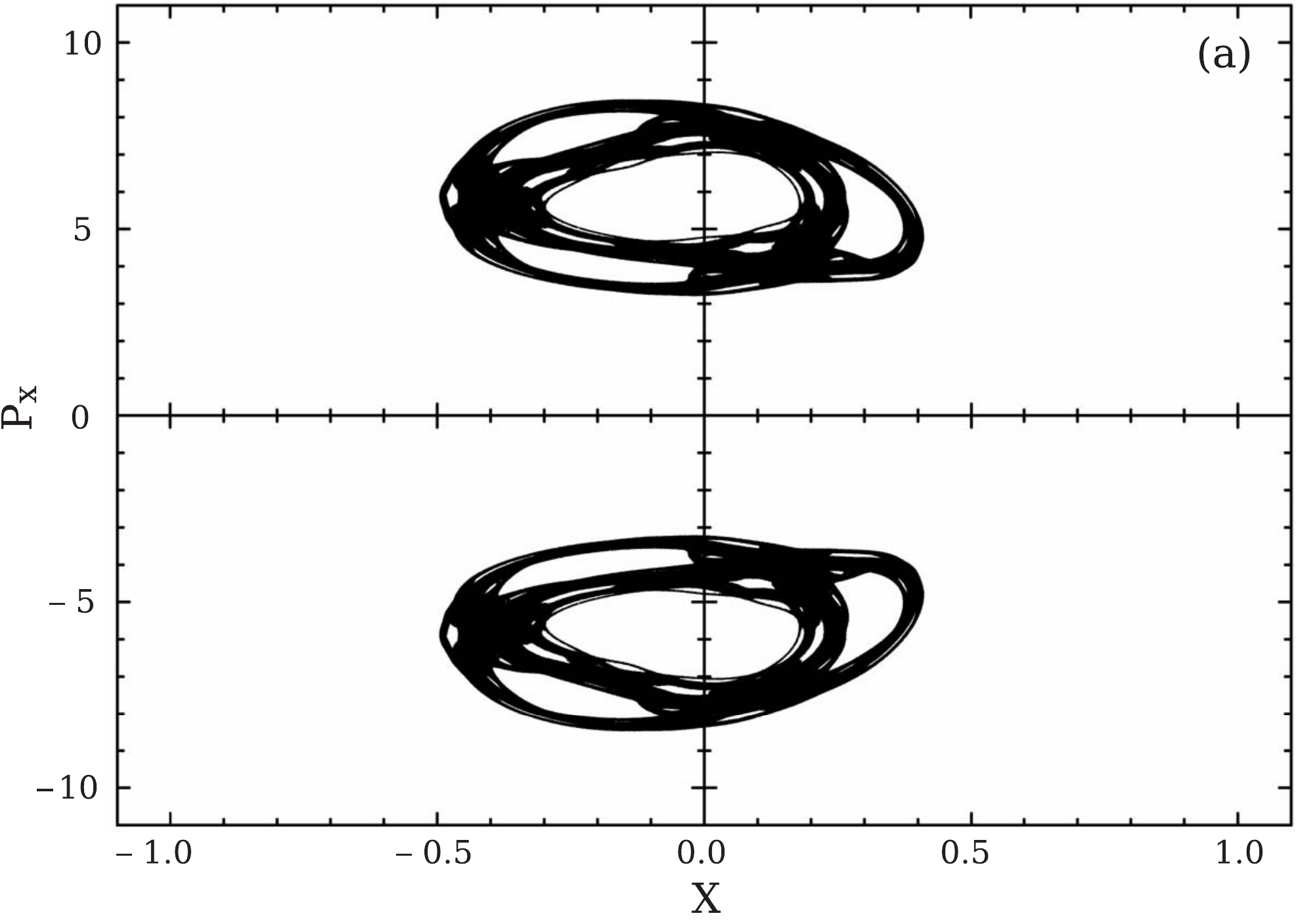}}\hspace{1cm}
                          \rotatebox{0}{\includegraphics*{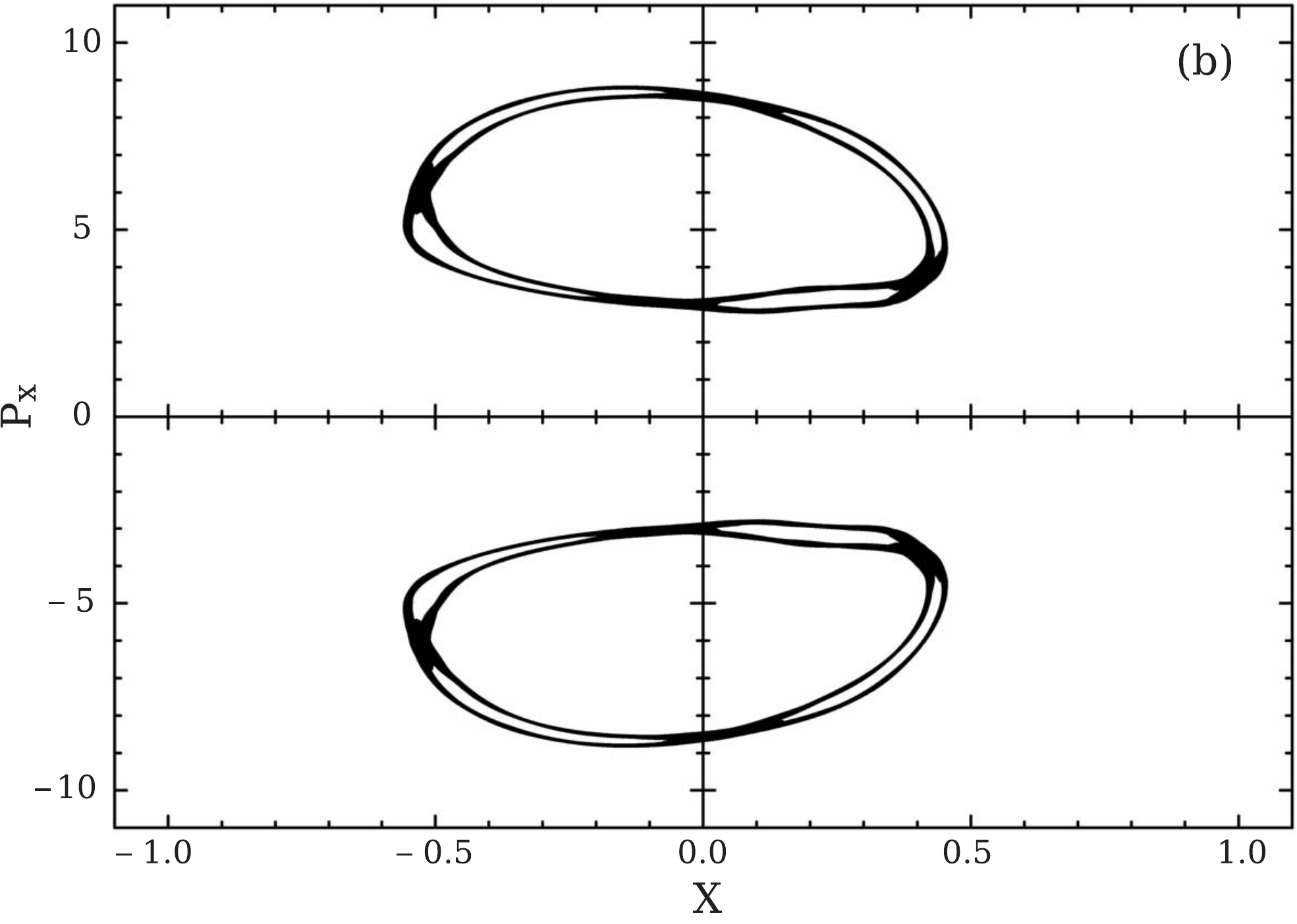}}}
\hbox{\includegraphics[width=60mm,angle=0]{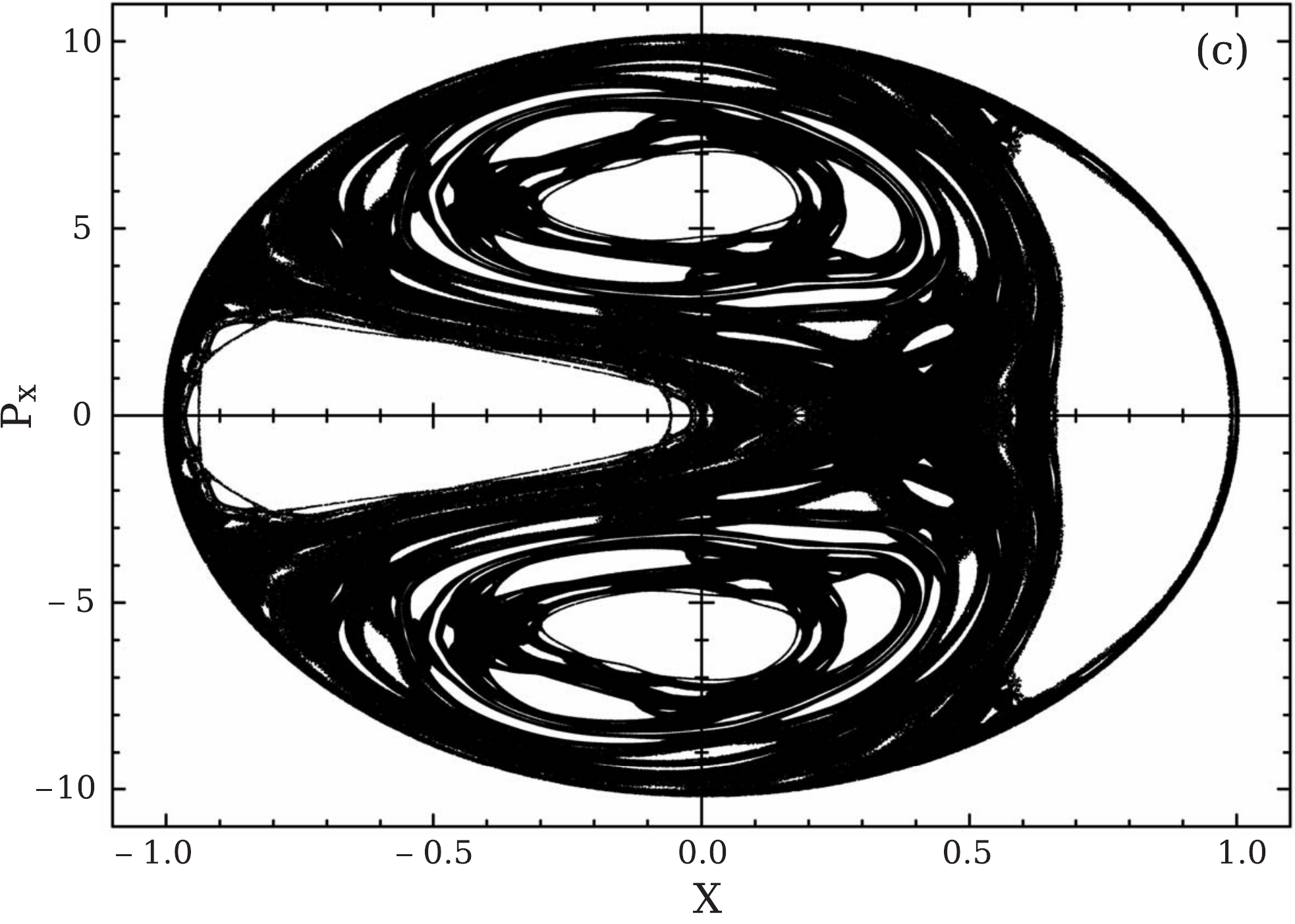}\hspace{0.3cm}
\includegraphics[width=60mm,angle=0]{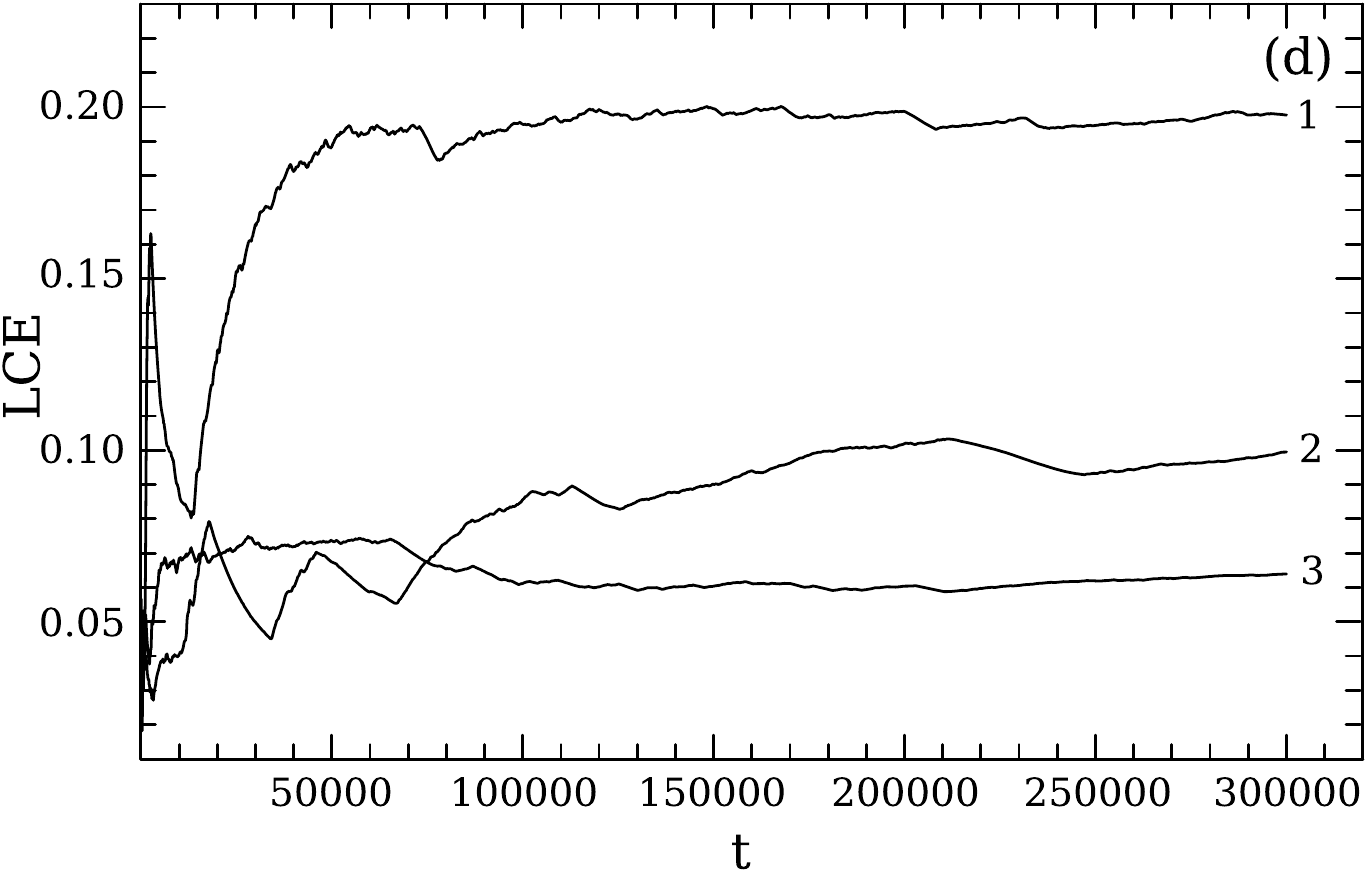}}
\end{center}
\captionb{6}{Panel (a):  chaotic component 2; panel (b):  chaotic component
3; panel (c):  all the chaotic components together; panel (d):  the LCEs
for the three chaotic components of the 2D system.}
\end{figure*}

\begin{figure}[!tH]
\centerline{\includegraphics[width=80mm,angle=0]{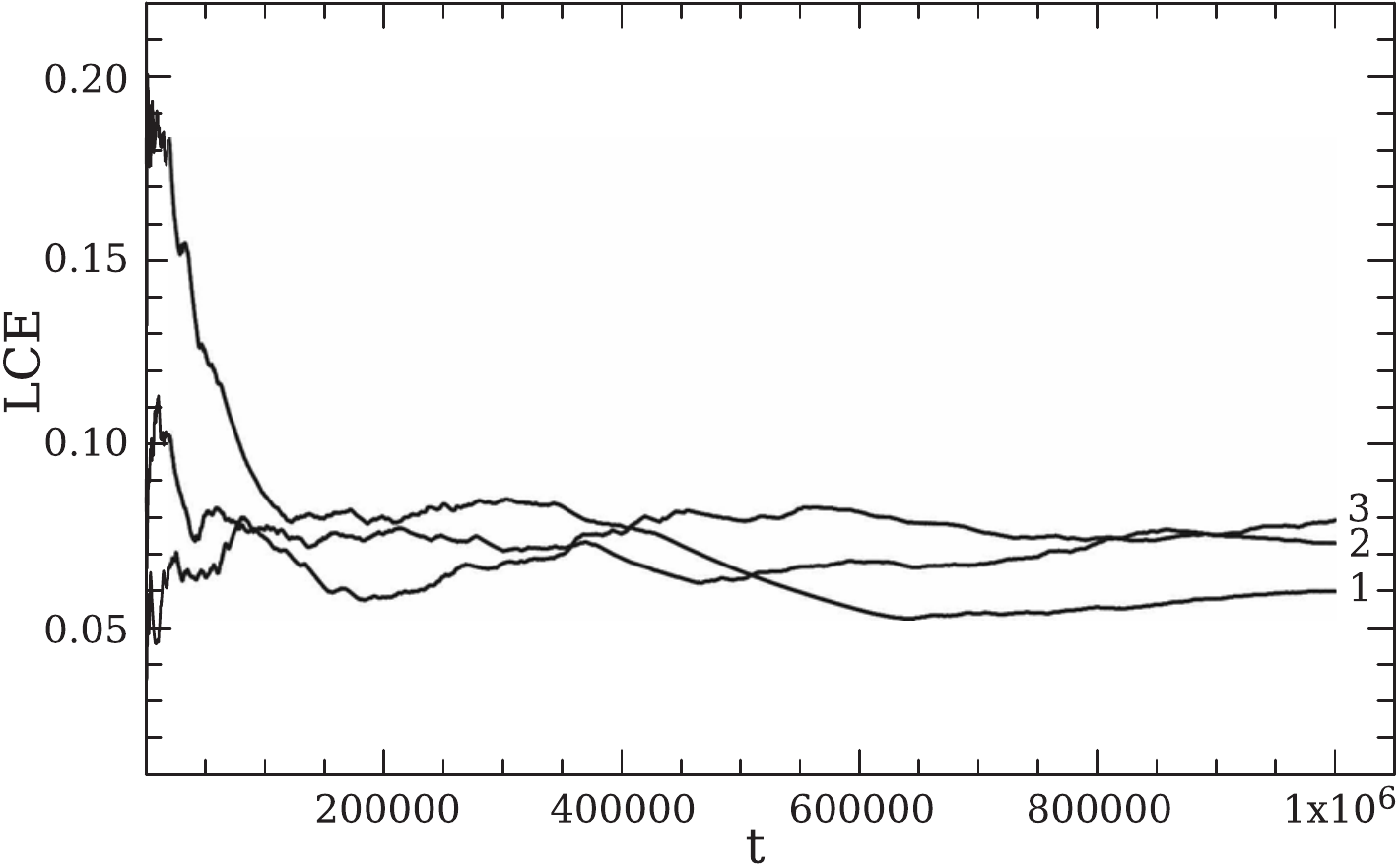}}
\hskip10mm{\captionb{7}{The LCEs for the three chaotic components of the 3D model.}}
\vskip5mm

\centerline{\hbox{\includegraphics[width=60mm,angle=0]{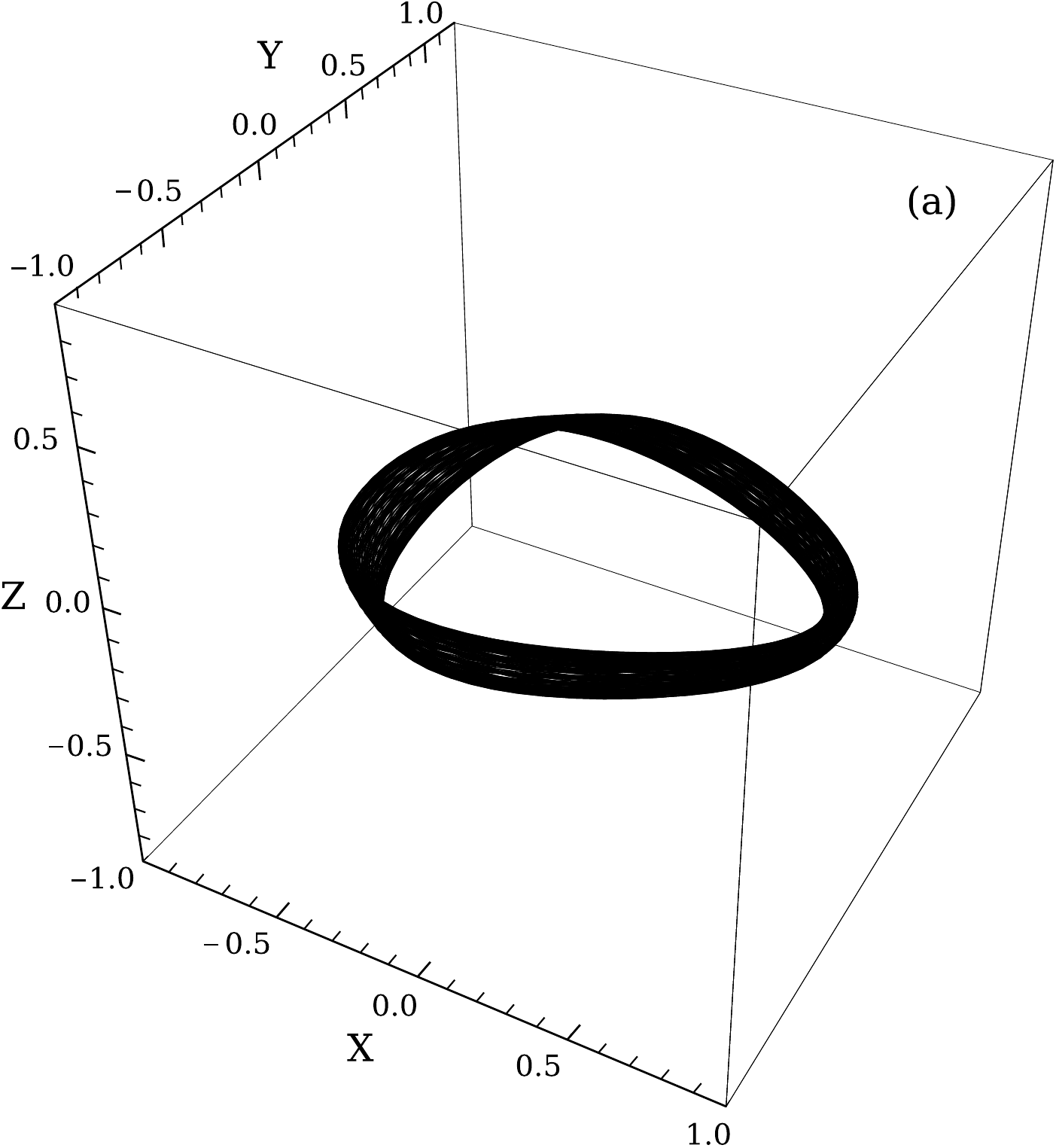}\hspace{0.3cm}
\raise1cm\hbox{\includegraphics[width=60mm,angle=0]{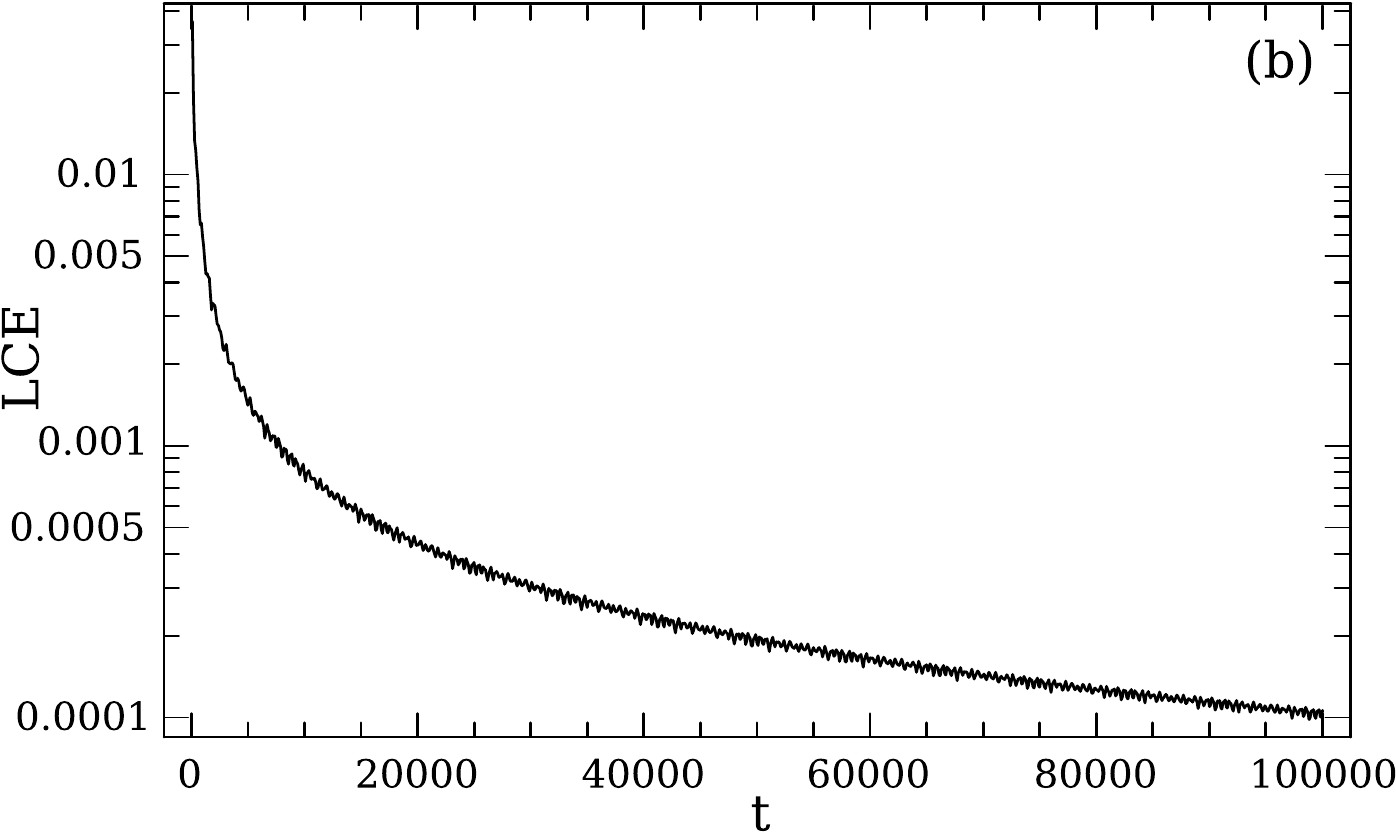}}}
}
\vskip3mm
\centerline{\hbox{\includegraphics[width=60mm,angle=0]{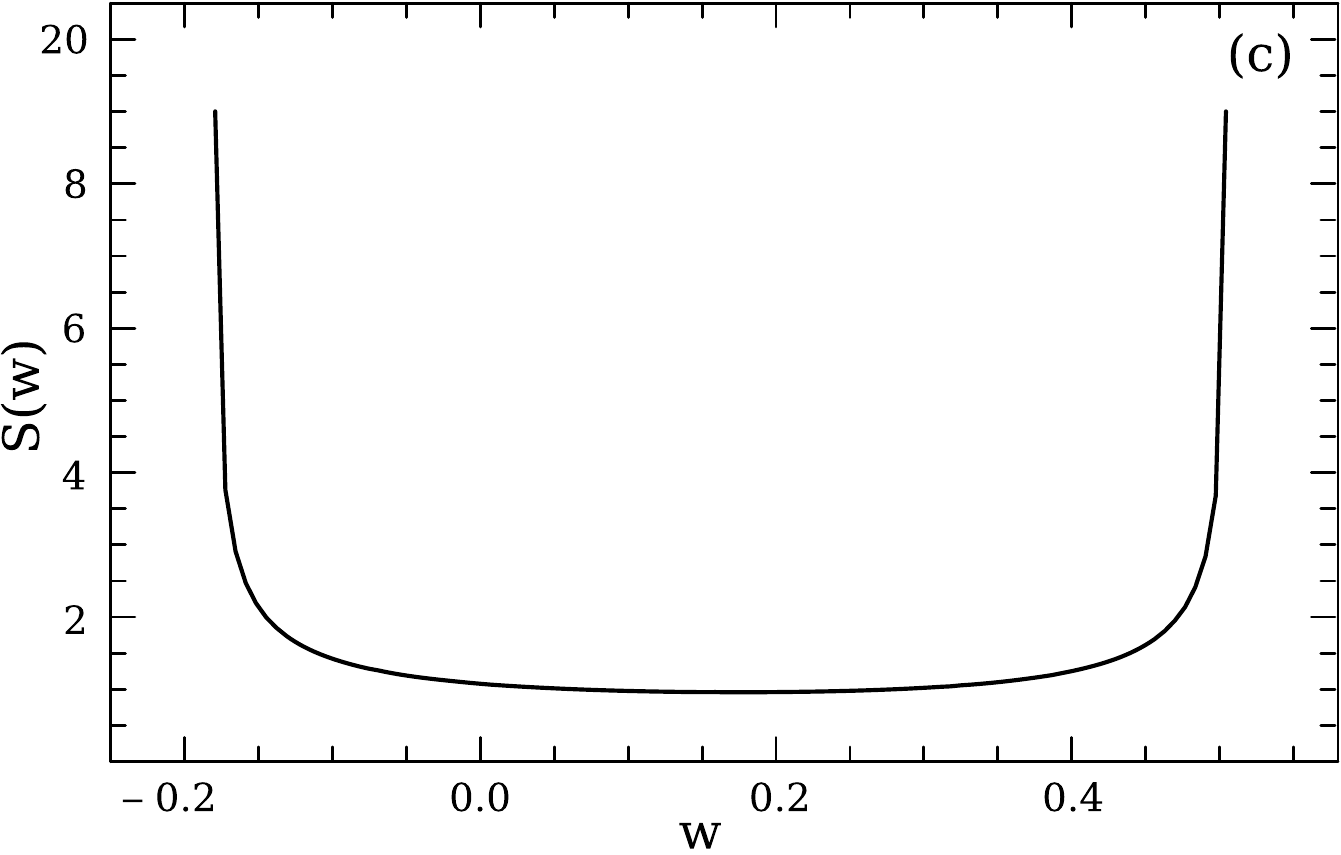}\hspace{0.3cm}
\raise1mm\hbox{\includegraphics[width=60mm,angle=0]{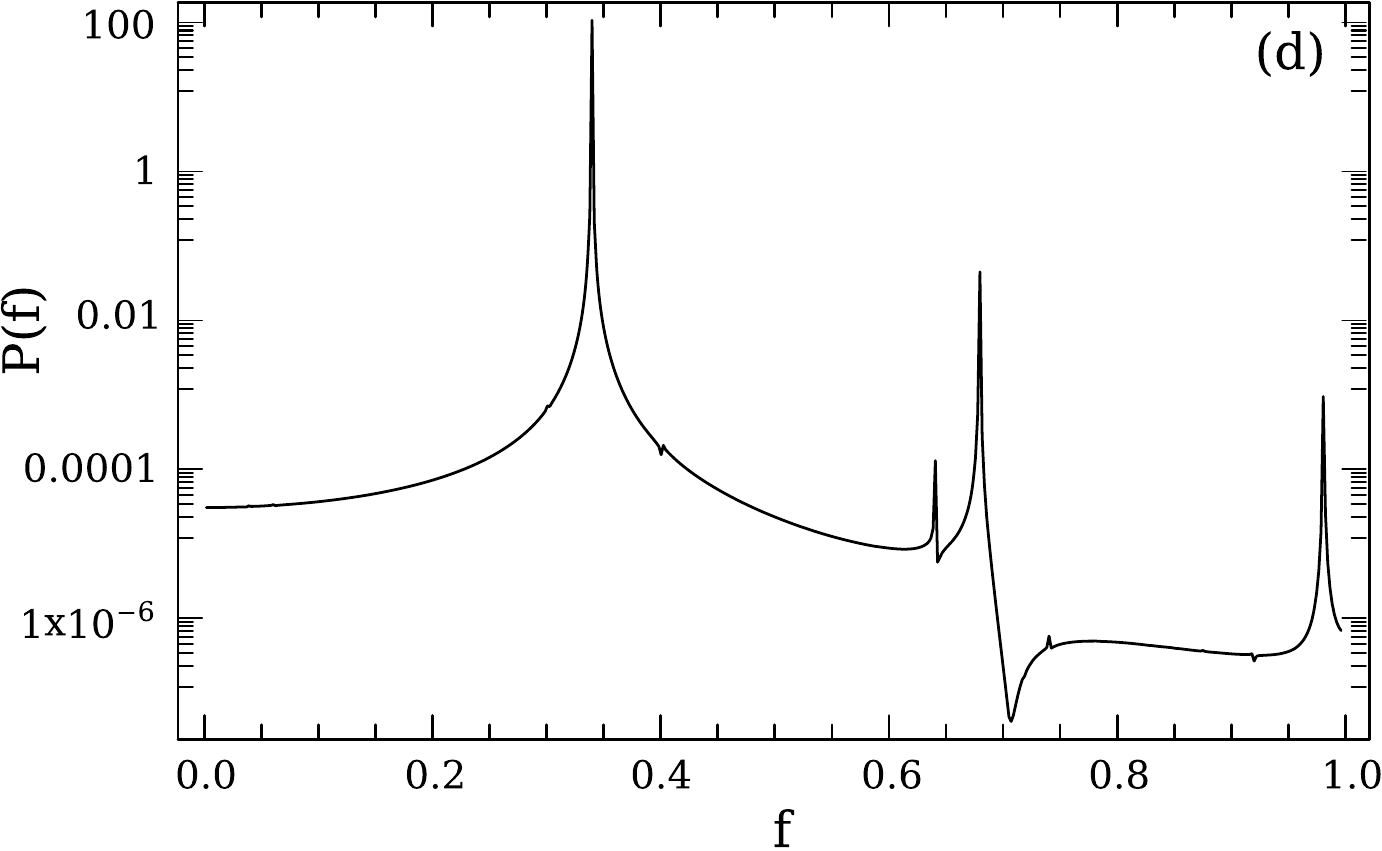}}}
}
\captionb{8}{Panel (a):  a 3D regular orbit; panel (b):  a plot of the LCE
vs. time for the orbit shown in (a); panel (c):  the $S(w)$ spectrum of
the orbit shown in (a); panel (d):  the $P(f)$ indicator for the orbit
shown in (a).  See the text for details.}
\end{figure}

The next step is to investigate different chaotic components of the 2D
system.  The corresponding results are shown in Figures 6 (a--d).
Figure 6a shows the chaotic region formed by an orbit with the initial
conditions:  $x_0=0.0$, $p_{x0}=4.6$ over a time period of $3\times
10^5$ time units.  This is the second chaotic component of the 2D
system.  Figure 6b shows the third chaotic component which is formed by
an orbit with the initial conditions:  $x_0=-0.51$, $p_{x0}=6.0$.  The
time period is $3\times 10^5$ time units.  Figure 6c shows all the
chaotic components together.  The LCEs of the above three chaotic
components are shown in Figure 6d.  The numbers 1,2 and 3 correspond to
the first, second and third chaotic component, respectively.  As we see,
each chaotic component has a different value of LCE (see Saito \&
Ichimura 1979).  Here we must note that a hierarchical structure in the
2D model is displayed not only by the stickiness but also by chaos.

\sectionb{3}{ORDER AND CHAOS IN THE 3D SYSTEM}

Now let we switch to study the behavior of the 3D system.  We take the
initial conditions $(x_0,p_{x0},z_0)$, $y_0=p_{z0}=0$, where
$(x_0,p_{x0})$ is a point on the phase planes of the 2D system.
Obviously this point lies inside the limiting curve
\begin{equation}
\frac{1}{2}p_x^2+V(x)=E_2 \ \ \ ,
\end{equation}
which is the curve
containing all the invariant curves of the 2D system.  We take $E=E_2$,
and the value of $p_{y0}$ for all orbits is obtained from the energy
integral (3).  Our numerical calculations in the 3D system show, that
the orbits with the
initial conditions ($x_0$, $p_{x0}$, $z_0$), $y_0 = p_{z0} = 0$, where
($x_0$, $p_{x0}$) is a point in the chaotic
regions of Figures 1 (a--d), for all permissible values of $z_0$
give
chaotic orbits.  On the other hand, it would be interesting to know what
happens with different chaotic components observed in the 2D system,
shown in Figures 6 (a--d).  The question is, do they merge to form a
unified
chaotic region in the 3D space, or they continue to exist as three
separate chaotic components in the 3D system?  In order to give an
answer, we computed the LCEs for three orbits, each starting in one of
the three different chaotic components, for a time period of $10^6$ time
units.  The results are shown in Figure 7. The numbers 1, 2 and 3
indicate the three chaotic components which have different LCEs values.
This result seems to be consistent with the outcomes obtained by
Cincotta et al.  (2006), where in a 3D system with divided phase space,
separate chaotic components actually exist.

One may ask an interesting question:  what is the nature of orbits which
have initial conditions ($x_0$, $p_{x0}$), $z_0$, $y_0 = p_{z0} = 0$,
where
($x_0$, $p_{x0}$) is a point in each regular region of Figures 1 (a--d)?
In order to give an answer, we will introduce and use a new type of
dynamical indicator,  the $S(w)$ spectrum.  The parameter $w_i$
is defined as
\begin{equation}
w_i=\frac{\left(x_i-p_{xi} \right) - \left(z_i-p_{zi} \right)}{p_{yi}} \ \ \ ,
\end{equation}
where ($x_i$, $z_i$, $p_{xi}$, $p_{yi}$, $p_{zi}$) are the
successive values of $(x, z, p_x, p_y, p_z)$ of the 3D orbit.  The
dynamical spectrum of the parameter $w$ is its distribution function
\begin{equation}
S(w)=\frac{\Delta N(w)}{N\Delta w} \ \ \ ,
\end{equation}
where $\Delta N(w)$ is the number of the parameters $w$ in the interval
$w, w+\Delta w$ after $N$ iterations.  In order to study the character
of a 3D orbit, the $S(c)$ spectrum can be also used. Note that the
coupling of the third component, $z$, carrying all the information about
the 3D motion, is hidden in the definition of the $S(c)$ spectrum, but
in any case it affects the values of $x, p_x$ and $p_y$.  Using the
definition of the $S(w)$ spectrum, we overtake this minor drawback and
create a new dynamical spectrum, suitable especially for 3D orbits.

\begin{figure*}[!t]
\centerline{\hbox{\includegraphics[width=60mm,angle=0]{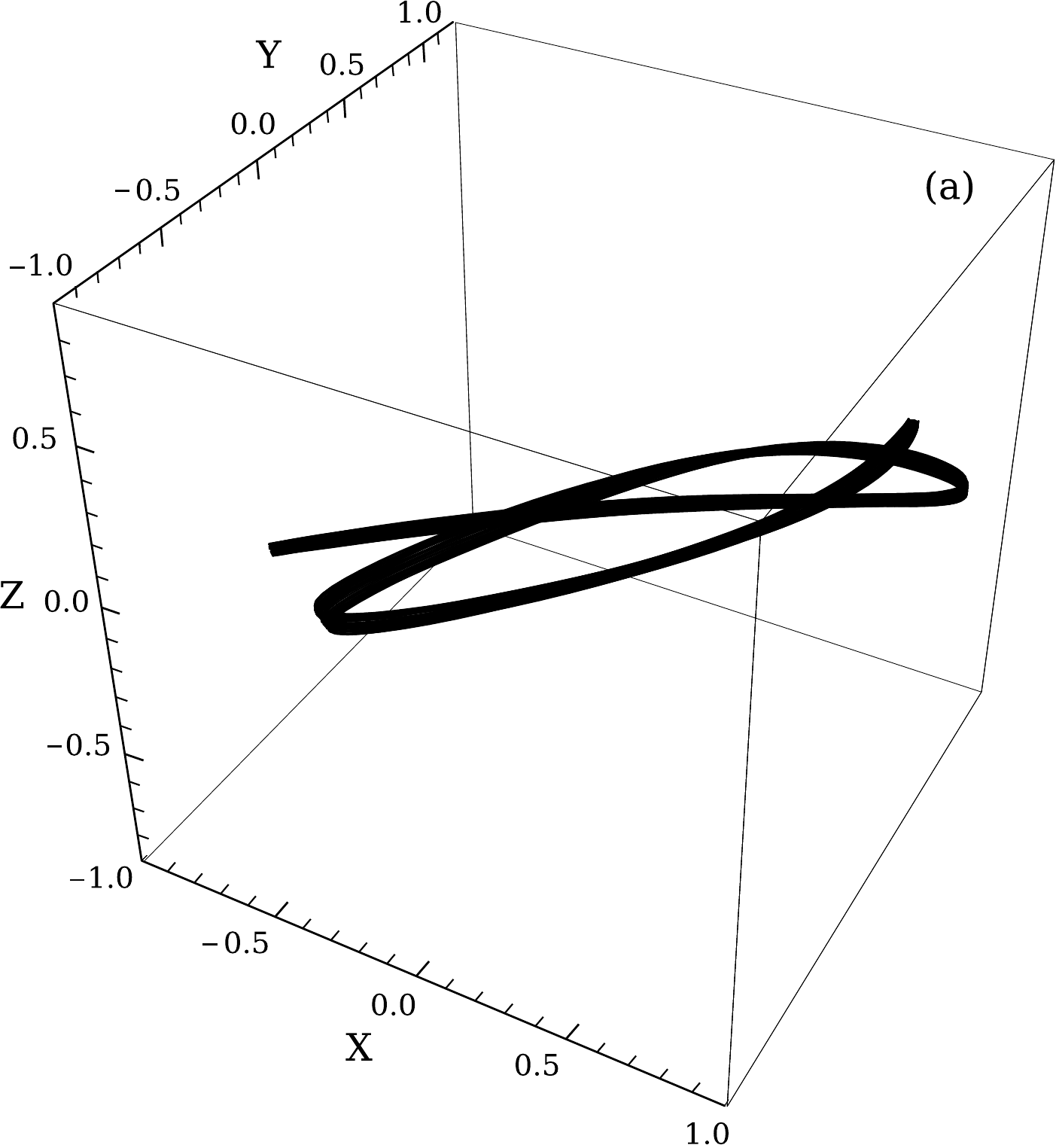}\hspace{0.3cm}
\raise1cm\hbox{\includegraphics[width=60mm,angle=0]{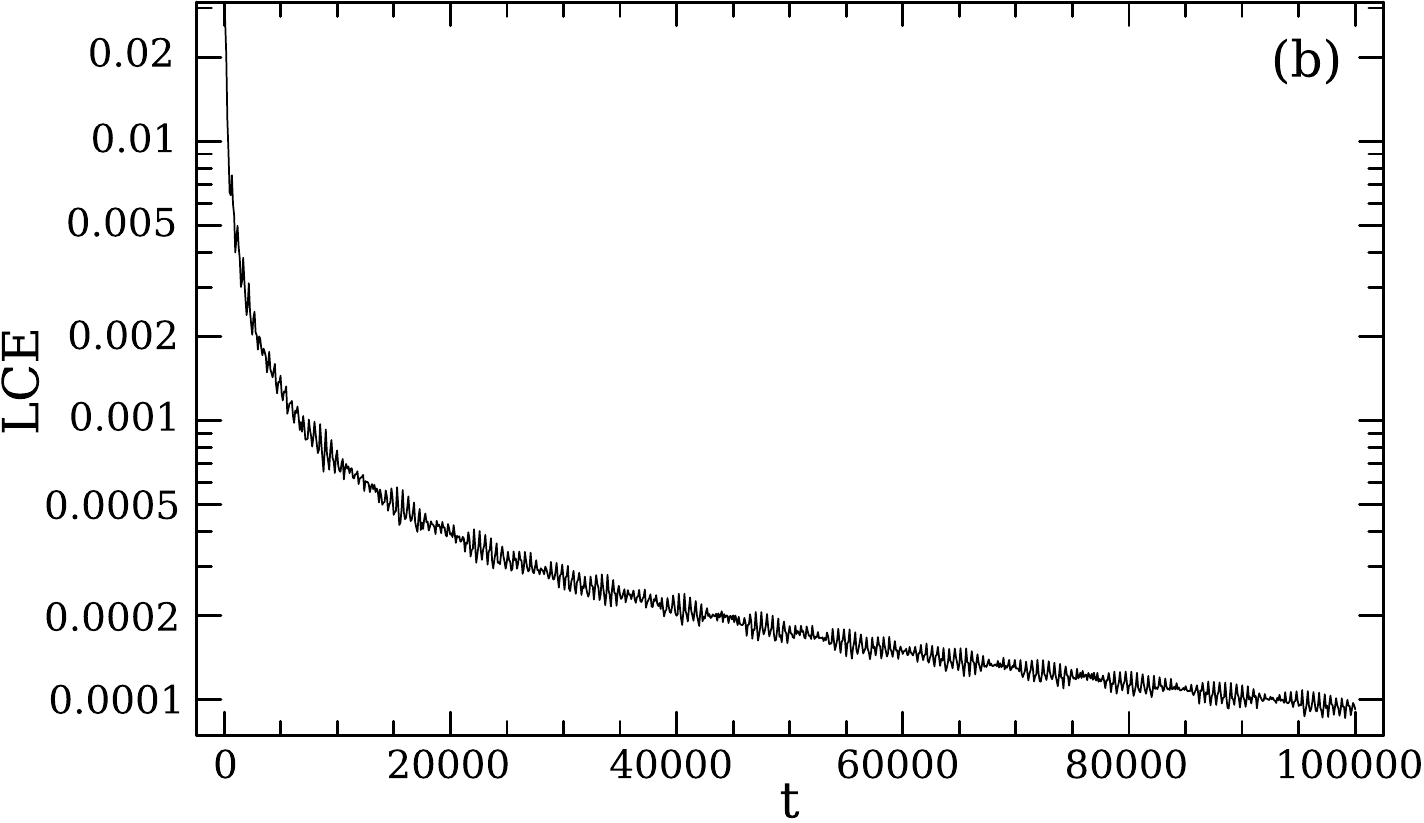}}}
}
\vskip3mm
\centerline{\hbox{\includegraphics[width=60mm,angle=0]{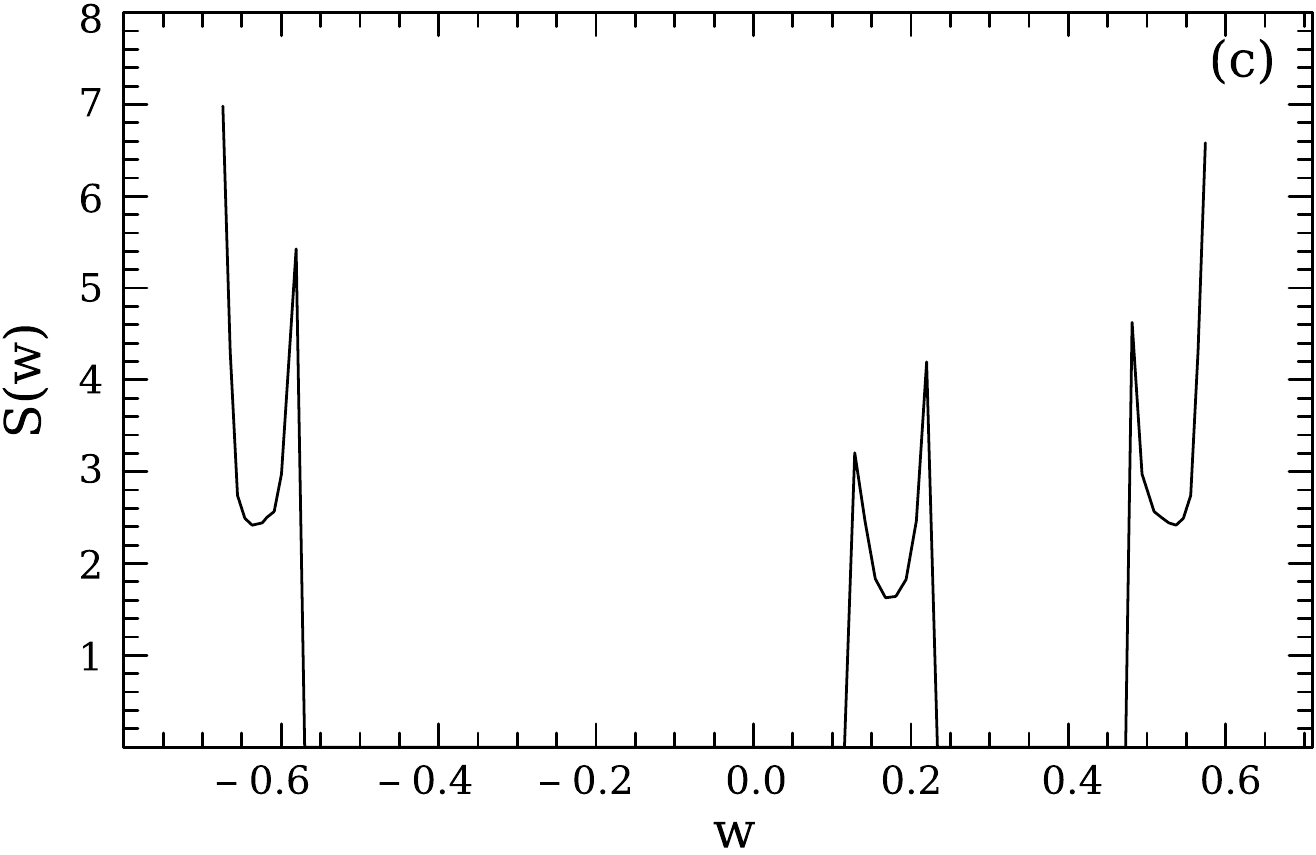}\hspace{0.3cm}
\raise1mm\hbox{\includegraphics[width=60mm,angle=0]{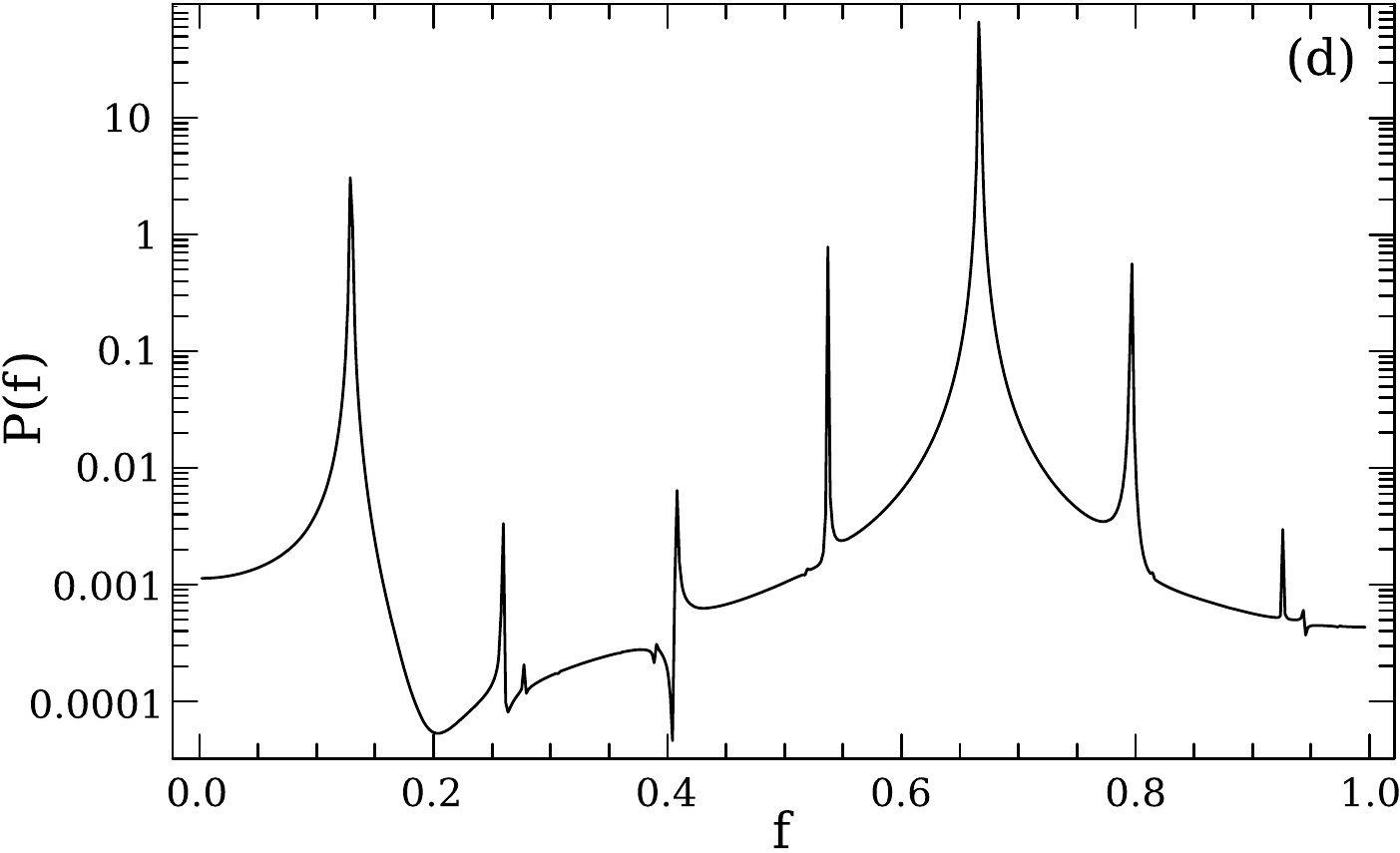}}}
}
\vskip-1mm
\captionb{9}{The same as in Figures 8 (a--d) but for a resonant 3D
orbit. See the text for details.}

\end{figure*}

\begin{figure*}[!t]
\begin{center}
\hbox{\includegraphics[width=60mm,angle=0]{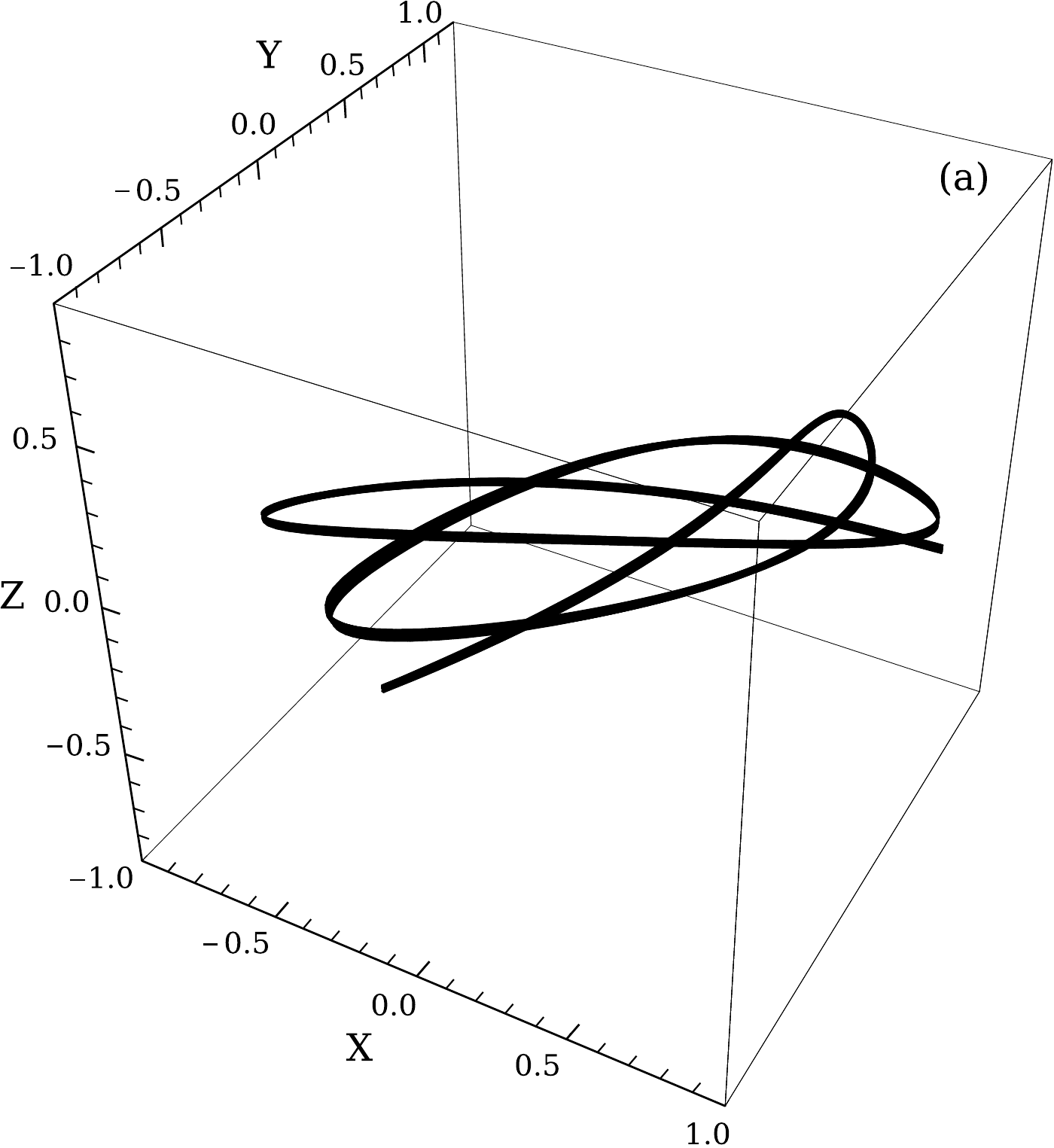}\hspace{0.3cm}
\raise1cm\hbox{\includegraphics[width=60mm,angle=0]{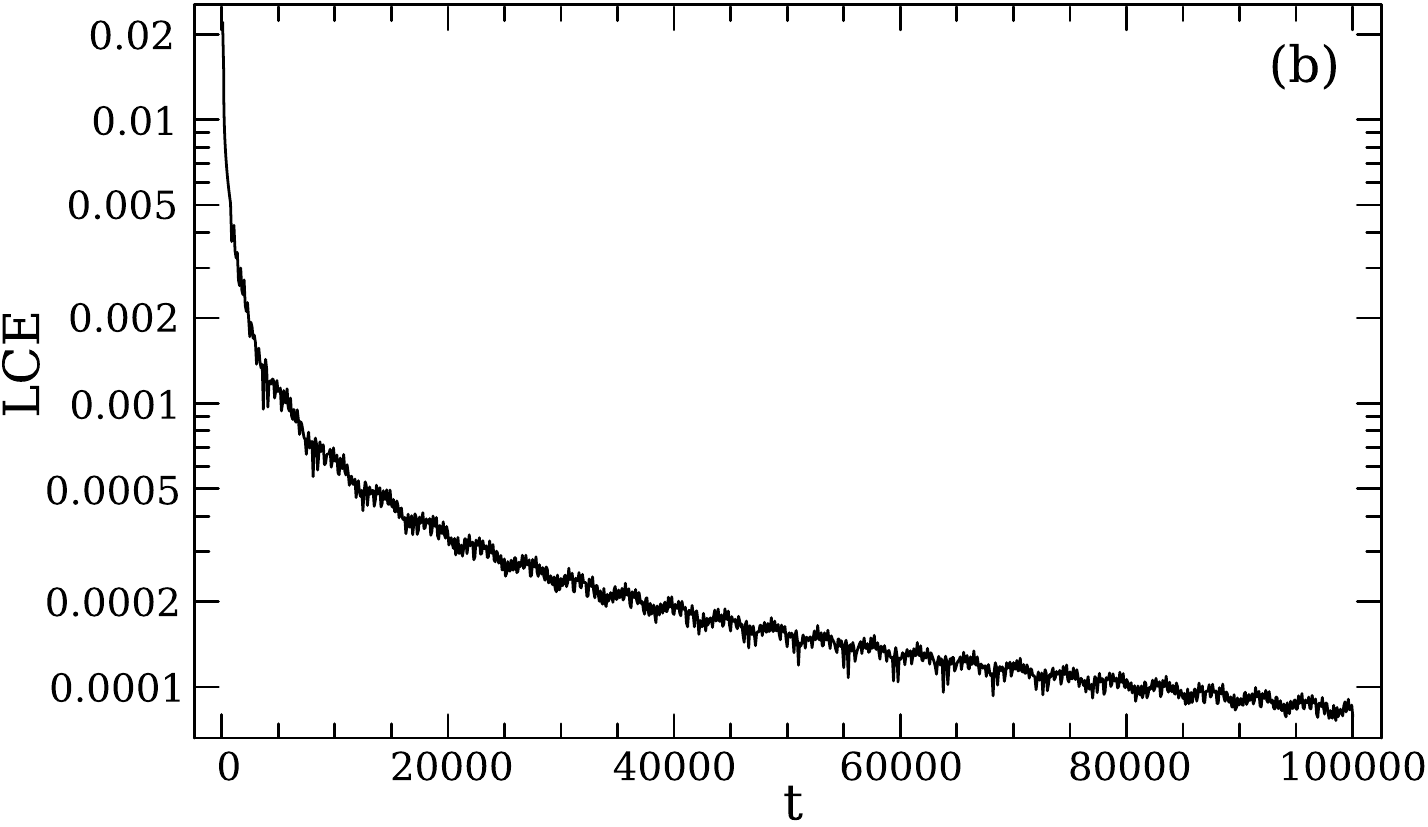}}}
\vskip3mm
\hbox{\includegraphics[width=60mm,angle=0]{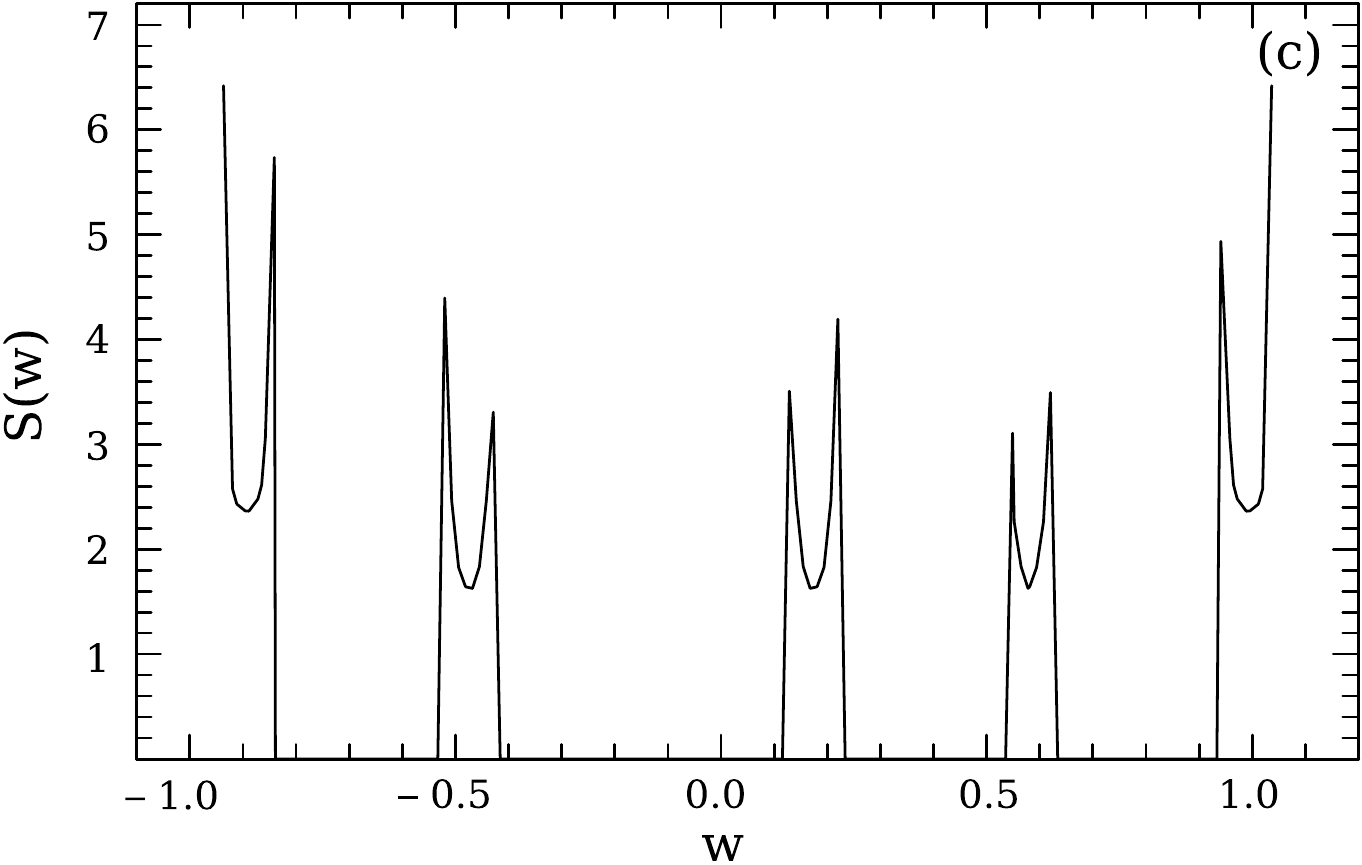}\hspace{0.3cm}
\raise1mm\hbox{\includegraphics[width=60mm,angle=0]{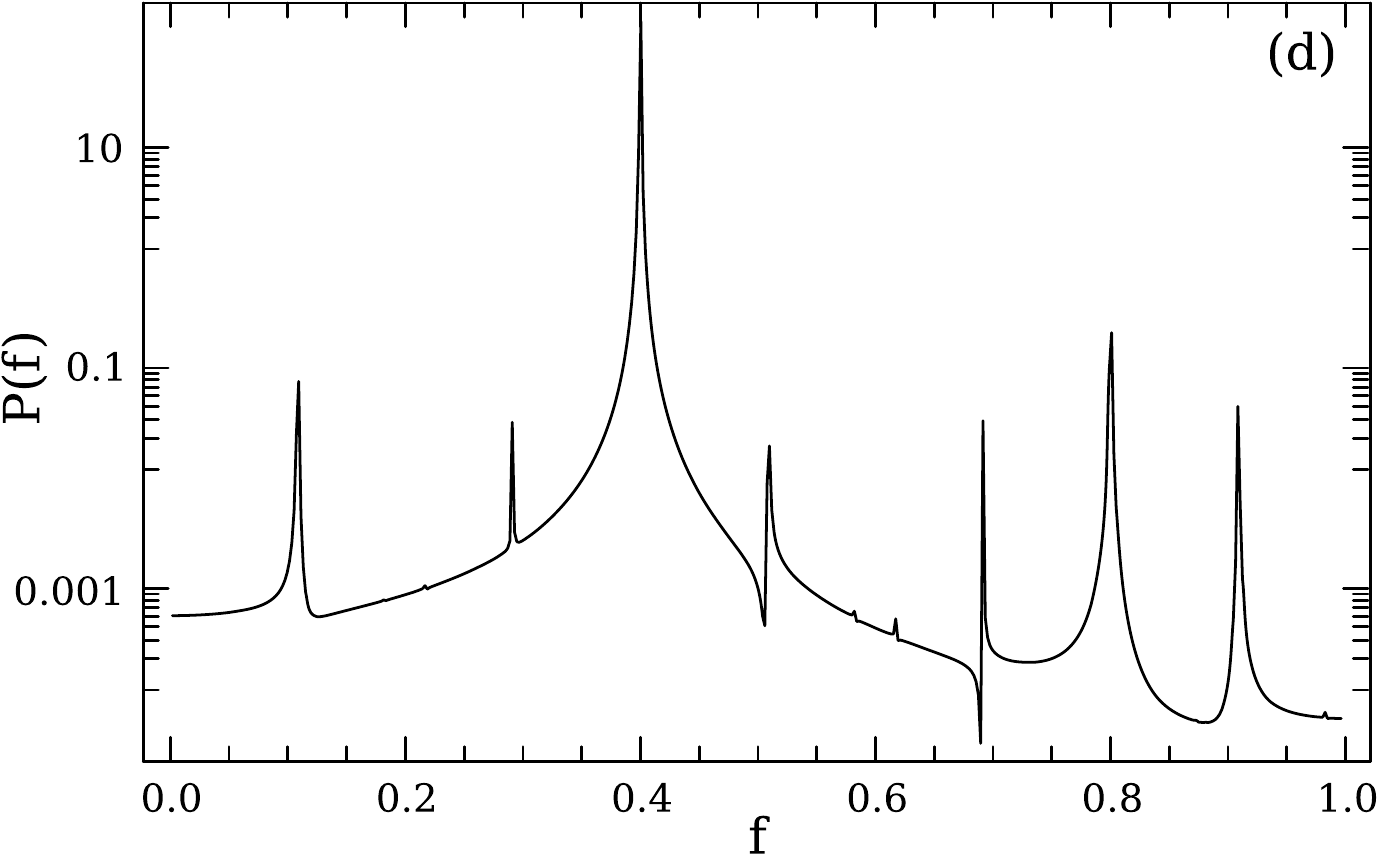}}}
\vspace{-7mm}
\end{center}
\captionb{10}{The same as in Figures 9 (a--d) but for a resonant 3D
orbit with different initial conditions.  See the text for details.}

\end{figure*}

\begin{figure*}[!t]
\begin{center}
\hbox{\includegraphics[width=60mm,angle=0]{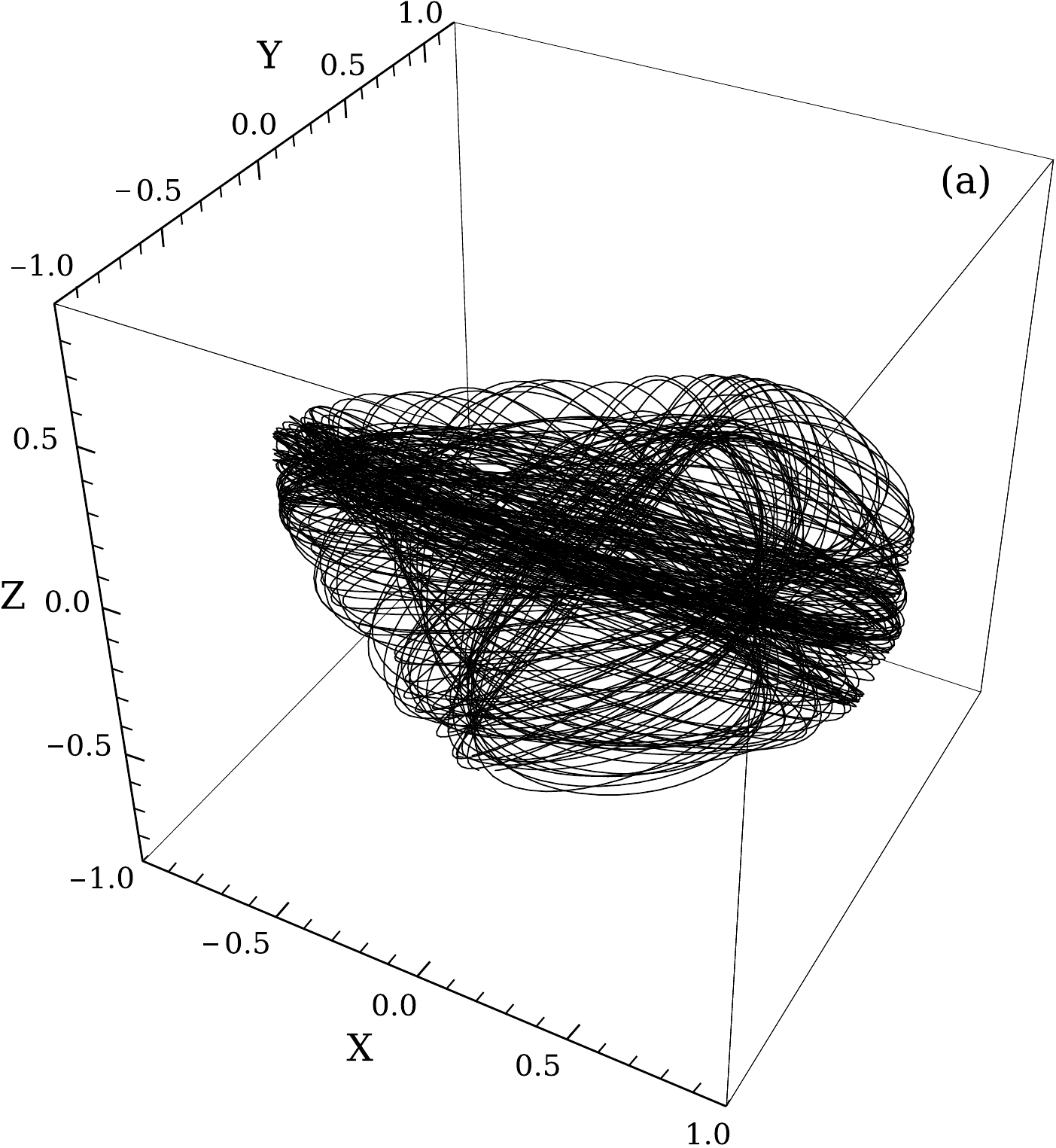}\hspace{0.3cm}
\raise1cm\hbox{\includegraphics[width=60mm,angle=0]{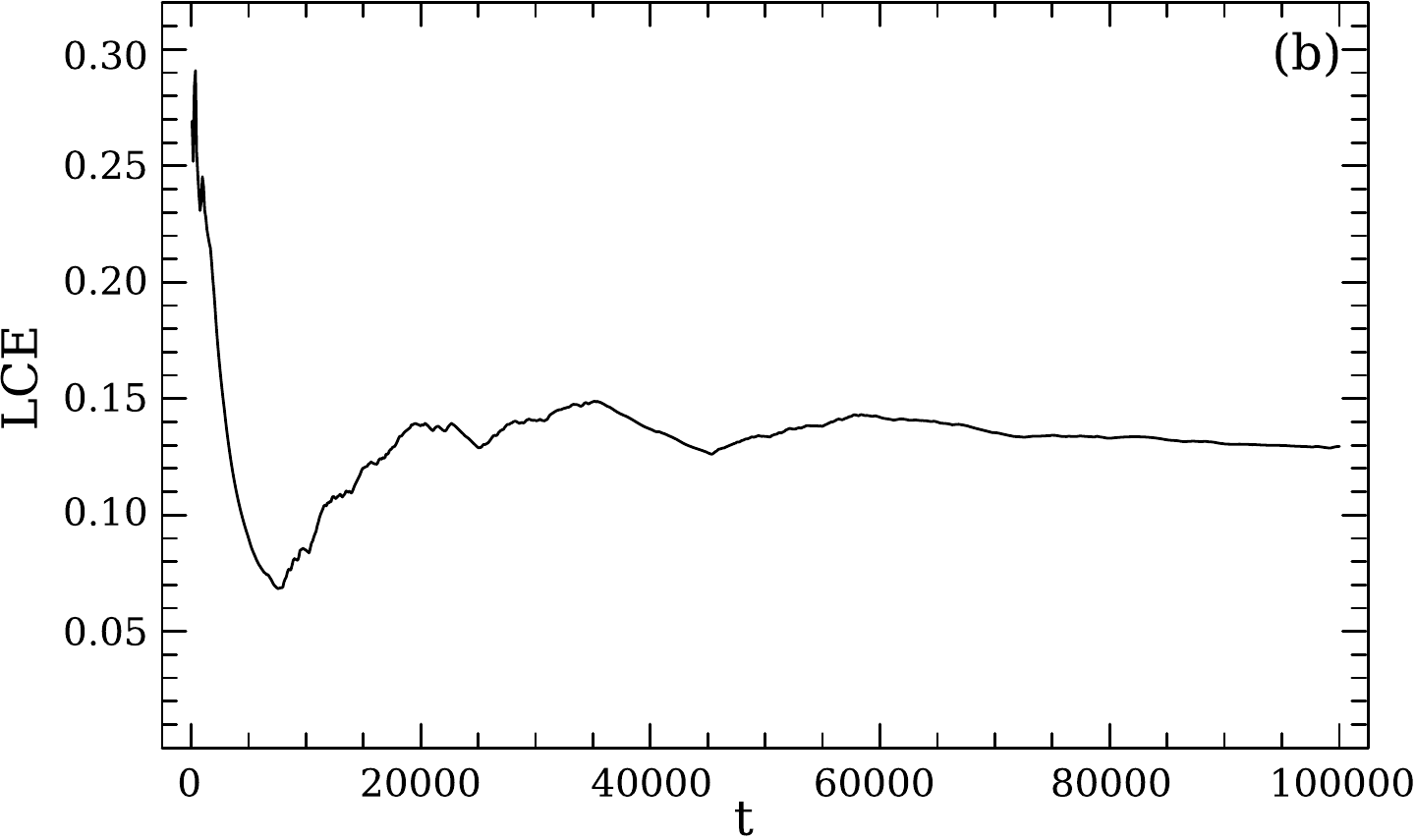}}}
\vskip3mm
\hbox{\includegraphics[width=60mm,angle=0]{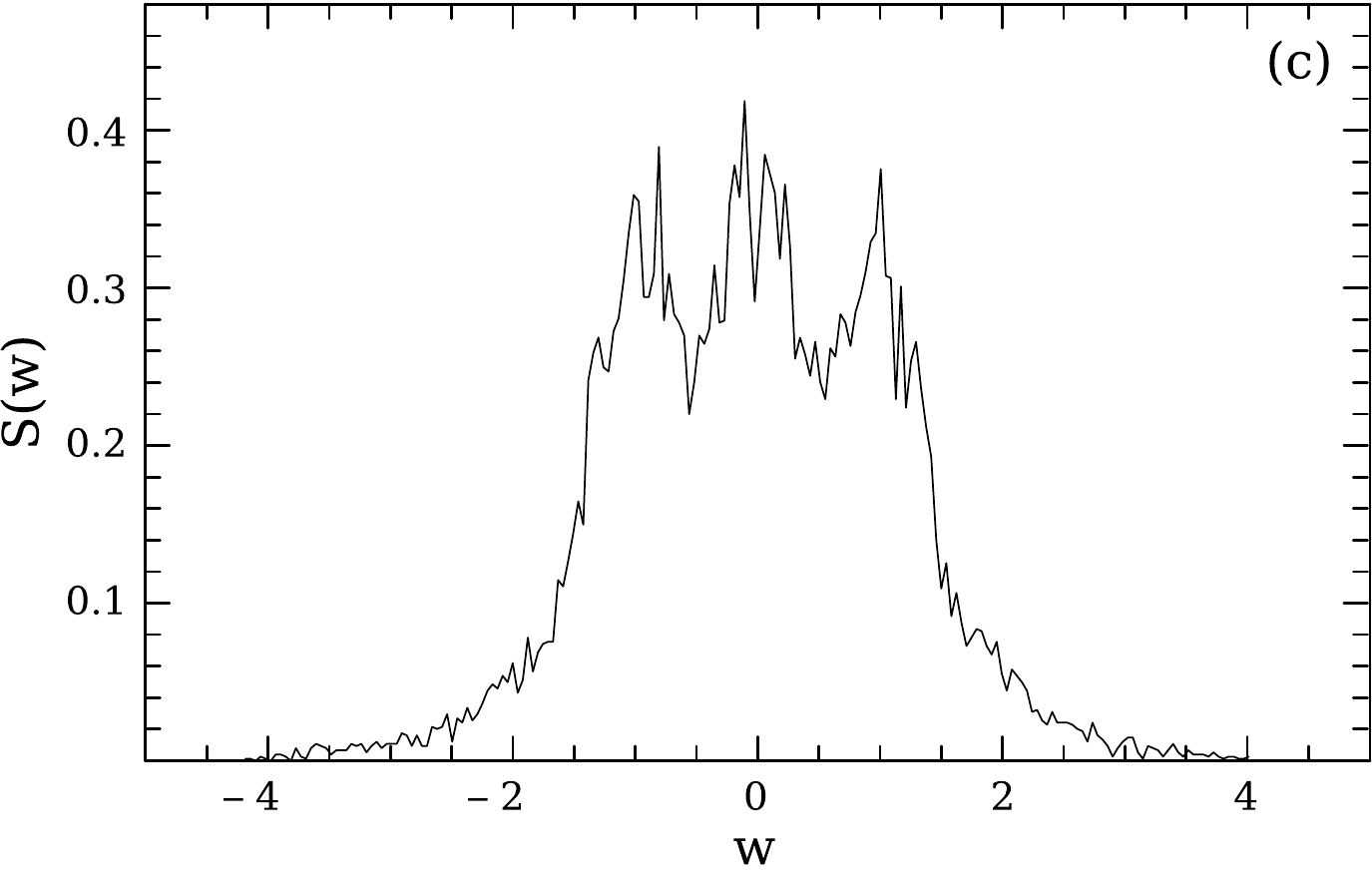}\hspace{0.3cm}
\raise1mm\hbox{\includegraphics[width=60mm,angle=0]{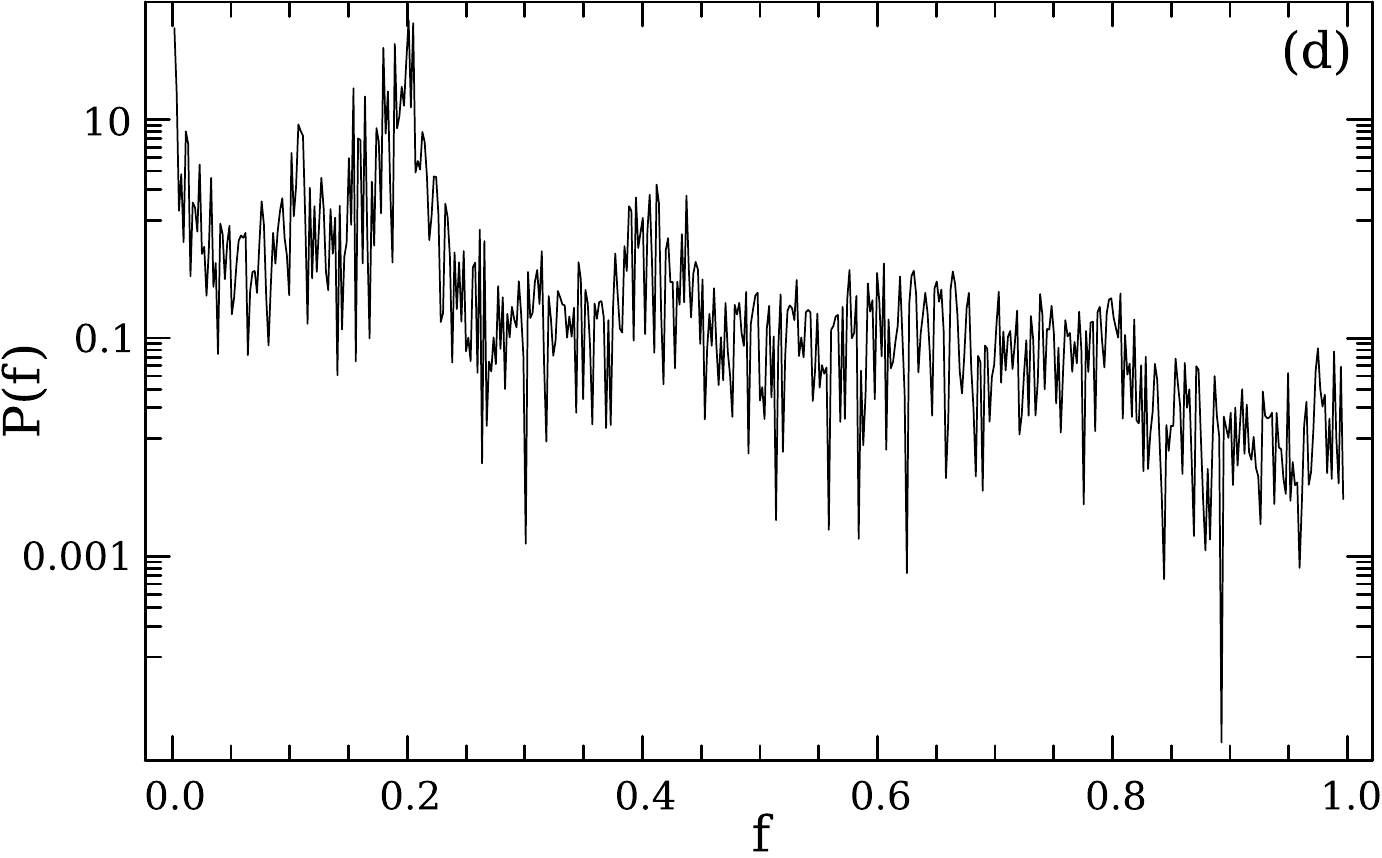}}}
\vspace{-7mm}
\end{center}
\captionb{11}{The same as in Figures 8 (a--d) but for a chaotic 3D
orbit. See the text for details.}
\end{figure*}

Figures 8 (a--d) show the results for a 3D regular orbit.  The orbit
which is shown in Figure 8a has the initial conditions:  $x_0=0.82$,
$y_0=p_{x0}=p_{z0}=0$, $z_0=0.01$, while for all 3D orbits the value of
$p_{y0}$ is always found from the energy integral (3).  The
corresponding values of all other parameters are as in Figure 1a.  The
LCE of the orbit, shown in Figure 8b, vanishes indicating the regular
motion.  Figure 8c shows the $S(w)$ spectrum of the orbit.  This is a
well defined $U$-type spectrum characteristic for the regular motion.
In Figure 8d, we see the $P(f)$ indicator which also indicates the
regular motion.  Figures 9 (a--d) are similar to Figures 8 (a--d) but
correspond to a 3D resonant orbit with the initial conditions:
$x_0=0.27$, $y_0=p_{z0}=0$, $p_{x0}=5.3$, $z_0=0.01$, producing three
$U$-type spectra.  The values of all other parameters are as in Figure
1b.

Figures 10 (a--d) are similar to Figures 9 (a--d) but for a 3D resonant
orbit with the initial conditions:  $x_0=0.47$, $y_0=p_{z0}=0$,
$p_{x0}=4.9$, $z_0=0.01$, producing five $U$-type spectra.  The values
of all other parameters are as in Figure 1b.  Finally, in Figure 11
(a--d), we present the results for a 3D chaotic orbit.  The initial
conditions are:  $x_0=0.09$, $y_0=p_{x0}=p_{z0}=0$, $z_0=0.1$.  The
values of all other parameters are as in Figure 1a.  Once more, we
observe that the results from all dynamical indicators, regarding the
character of motion, coincide.  The integration time for all 3D orbits
is 100 time units, for the $S(w)$ spectrum it is $2 \times 10^4$ time
units and for the $P(f)$ indicator it is $4 \times 10^3$ time units.

Using the above method, we have computed a large number of orbits (about
1000) in the 3D dynamical system.  Numerical outcomes suggest that for
all ($x_0$, $p_{x0}$) in the regular regions of Figures 1 (a--d) and
for small values of $z_0$ ($z_0$\,\dnnn {$\stackrel{\textstyle
<}{\sim}$}\,0.12) the motion is regular, while for all larger
permissible values of $z_0$ the motion becomes chaotic.

\sectionb{4}{DISCUSSION AND CONCLUSIONS}

In this article we have studied the properties of motion in a 3D
Hamiltonian system, describing the motion in the inner parts of a
deformed galactic model.  We started our investigation from the 2D
system, because orbits confined in the galactic plane $(z= 0)$ display
some very interesting features, such as sticky regions, chaotic
components and islandic motion produced by a large number of secondary
resonances.

Results from the study of the phase planes indicate, that there is an
hierarchy in sticky regions.  A test particle can stay in sticky regions
for a time period of about 75\,000 time units, before leaving to the
corresponding chaotic region.  Several chaotic components are also
observed in the 2D system, each one having a different value of LCE.
Thus we conclude, that both the stickiness and chaos display a
hierarchical structure in the 2D model.  An interesting result of this
investigation is that the percentage $A\%$ of the surface of the section
occupied by chaotic orbits decreases, tending asymptotically to zero,
when the mass of the disk increases.  This suggests that disks in
elliptical galaxies can act as the chaos controllers.

Our numerical calculations in the 3D system show, that orbits with the
initial conditions ($x_0$, $p_{x0}$, $z_0$), $y_0=p_{z0}=0$, where
($x_0$, $p_{x0}$) is a point in the chaotic regions of Figure 1 (a--d),
for all permissible values of $z_0$ give chaotic orbits.  Using the new
$S(w)$ spectrum we find that orbits with the initial conditions ($x_0$,
$p_{x0}$, $z_0$), $y_0=p_{z0}=0$, where ($x_0$, $p_{x0}$) is a point in
the regular regions of Figure 1 (a--d), for small values of $z_0$ are
regular, while for larger values of $z_0$ they become chaotic.  It is
also of particular interest that the 3D system displays three different
chaotic components, not a unified chaotic region.

A very effective and reliable tool to distinguish between the regular
and chaotic motion in the 3D dynamical systems is the new $S(w)$
spectrum.  This spectrum, which is an advanced form of the $S(c)$
spectrum, allows us to detect islandic 3D motion of the resonant orbits,
since it produces as much spectra as the total number of islands in the
$x-p_x$, $y=0$, $p_y>0$ surface of section (see Figures 9a and 10a).
Moreover, the comparison with other dynamical parameters, such as the
LCE and the $P(f)$ indicator, shows that the results obtained by the
$S(w)$ spectrum are reliable.

\thanks{ The author would like to thank Prof.  N. D. Cara\-nicolas for
his fruitful discussions during this research.  I also would like to
thank Peeter Tenjes for careful reading of the manuscript and useful
comments which allowed us to improve the quality of the paper.}

\References
\parskip=-0.4pt

\refb Barazza F. D., Binggeli B., Jerjen H. 2002, A\&A, 391, 823

\refb Bingelli B., Tammann G. A., Sandage A. 1987, AJ, 94, 251

\refb Caranicolas N. D., Papadopoulos N. 2007, AN, 328, 556, 259

\refb Caranicolas N. D., Zotos E. E. 2010, New Astronomy, 15, 427

\refb Cincotta P. M., Giordano C. M., Perez M. J. 2006, A\&A, 455, 499

\refb Conselice C. J., Gallagher J. S. III, Wyse R.\,F.\,G. 2001, ApJ,
559, 791

\refb Contopoulos G., Polymilis C. 1993, Phys.  Rev.  E, 46(3), 1546

\refb Contopoulos G., Varvoglis H., Barbanis B. 1987, A\&A, 172, 55

\refb De Rijcke S., Dejonghe H., Zeilinger W. W., Hau G.\,K.\,T. 2001,
ApJ, 559, 21

\refb Feruson H. C., Sandage A. 1989, ApJ, 346, 53

\refb Jerjen H., Kalnajs A., Binggeli B. 2000, A\&A, 358, 845

\refb Karanis G. I., Caranicolas N. D. 2002, AN, 323, 3

\refb Karanis G. I., Vozikis Ch.  L. 2008, AN, 329, 403

\refb Khochfar S., Burkert A. 2005, MNRAS, 359, 1379

\refb Lichtenberg A. J., Lieberman M. A. 1992, {\it Regular and
Chaotic Dynamics}, 2nd Edition, Springer

\refb Mayer L., Governato F., Colpi M. et al. 2001, ApJ, 547, 123

\refb Merrifield M. R., Kuijken K. 1999, A\&A, 345, L47

\refb Miyamoto W., Nagai R. 1975, PASJ, 27, 533

\refb Moore B., Lake G., Katz N. 1998, ApJ, 495, 139

\refb Ryden B. S., Terndrup D. M., Pogge R. W. 1999, ApJ, 517, 650

\refb Saito N., Ichimura A. 1979, in {\it Stochastic Behaviour in
Classical and Quantum Hamiltonian Systems}, eds.  G. Casati \& L. Ford,
Springer, p.\,137

\refb Simien F., Prugniel Ph. 2002, A\&A, 384, 371

\refb Siopis C., Contopoulos G., Kandrup H. E. 1995, NYASA, 751, 205

\refb Young L. M. 2002, AJ, 124, 788

\refb Young L. M. 2005, ApJ, 634, 258

\end{document}